\documentclass[runningheads,orivec]{llncs}
\usepackage[T1]{fontenc}
\usepackage{enumitem}
\usepackage{booktabs, tabularx}
\usepackage{array}
\usepackage{makecell}
\usepackage{amsmath,amsfonts}
\usepackage{algorithmic}
\usepackage[caption=false,font=normalsize,labelfont=sf,textfont=sf]{subfig}
\usepackage{textcomp}
\usepackage{stfloats}
\usepackage{url}
\usepackage{verbatim}
\usepackage{graphicx}
\usepackage{xspace}
\usepackage{pifont}
\usepackage{bm}
\usepackage{threeparttable}
\usepackage{multirow}
\usepackage{xcolor}
\usepackage[most]{tcolorbox}
\usepackage{comment}
\usepackage{hyperref}
\usepackage{cleveref}
\usepackage{siunitx}
\usepackage{bbding}
\usepackage{tikz}

\begin{document}

\title{AndroTruth: A Reliable Benchmark Android Malware Dataset Derived from Technical Expert Reports}

\titlerunning{AndroTruth}

\author{
Hongpeng Bai\inst{1}\thanks{Hongpeng Bai and Yao Zhang contributed equally to this work.}
\and
Yao Zhang\inst{1}\inst{\star}
\and
Minhong Dong\inst{1}
\and
Shunzhe Zhao\inst{2}
\and
Haobo Zhang\inst{2}
\and
Lingyue Li\inst{1}
\and
Yude Bai\inst{3}\thanks{Yude Bai and Guangquan Xu are corresponding authors.}
\and
Shuai Hu\inst{4}
\and
Guangquan Xu\inst{1}\inst{\star\star}
}

\authorrunning{H. Bai et al.}

\institute{
School of Cyber Security, Tianjin University, Tianjin, China\\
\email{\{bai931214,zzyy,dongminhong,lingyuel,losin\}@tju.edu.cn}
\and
School of Cyber Science and Technology, Shihezi University, Shihezi, Xinjiang, China\\
\email{\{zhaoshunzhe,zhanghaobo\}@stu.shzu.edu.cn}
\and
School of Software, Tiangong University, Tianjin, China\\
\email{baiyude@tiangong.edu.cn}
\and
State Grid Xinjiang Electric Power Research Institute, Xinjiang, China
\email{hushuai543@163.com}
}

\maketitle
\begin{abstract}
Reliable family labels are essential for Android malware analysis, yet most widely used benchmarks derive such labels from aggregated VirusTotal engine outputs. Because antivirus vendors differ in detection logic, naming conventions, and signature updates, these labels are often inconsistent across engines and unstable over time, which introduces substantial noise into downstream empirical evaluation.

To address this problem, we construct \textit{AndroTruth}, an Android malware family benchmark whose labels are derived exclusively from traceable expert technical analysis reports rather than AV-consensus voting. AndroTruth spans 2016 to 2025 and contains 8,172 malware samples from 187 families. Our statistical results show that automated labeling tools can exhibit a misleading consensus failure mode in which AVClass2 and ClarAVy agree with each other yet jointly disagree with expert ground truth on 25.38\% of samples with explicit labels from both tools. 

Experimental results show that, under expert-verified supervision, representative classifiers such as Meta-MAMC and AndMFC achieve accuracy above 96\%. When trained with real-world AV-derived labels and evaluated against expert ground truth, however, their performance drops to
only about 60\% accuracy and about 35\% macro-F1. ClarAVy confidenceaware filtering can improve family grouping quality, but cannot replace expert-verified labels for exact family naming. Together, these results demonstrate that label reliability is a first-order factor in Android malware family evaluation and highlight the need for expert-verified benchmarks.
\end{abstract}




\section{Introduction}\label{sec:intro}

Detecting Android malware typically relies on labeled benchmark datasets for model training and empirical evaluation. Most widely used datasets, from early benchmarks such as Drebin to more recent resources such as BCG, derive malware family labels from VirusTotal (VT) by aggregating scan results from multiple antivirus (AV) engines. Although this strategy offers scale and convenience, it raises persistent concerns about label reliability~\cite{10.1145/3618257.3624800,10.1145/3691620.3695280,zhu2020measuring}.

Antivirus engines differ in detection strategies, signature maintenance, and naming conventions~\cite{zhu2020measuring,10.1145/3618257.3624800}, which can produce conflicting family assignments for the same sample. VirusTotal derived labels may also drift over time as vendors update signatures and taxonomies. For example, re-scanning samples from the Drebin dataset reveals that about 19\% of family labels differ from the original Kaspersky annotations~\cite{DBLP:conf/ndss/ArpSHGR14,spreitzenbarth2013mobile}. Automatic labeling tools such as AVClass~\cite{sebastian2016avclass,sebastian2020avclass2} and ClarAVy~\cite{joyce2025claravy} can reduce naming inconsistency, but they still aggregate engine outputs rather than consult verified ground truth. Consequently, they may propagate structured family-label errors when multiple engines make the same mistake.

Unreliable family labels can distort studies of malware evolution, family relationships, and variant analysis~\cite{bai2023argusdroid,10.1145/3719027.3765197,10179453}, while evaluation results derived from such data may become unstable or misleading, especially under long-tail family distributions and temporally shifted test settings~\cite{bai2020unsuccessful}. Existing Android malware benchmarks therefore face two related limitations. First, aggregation-based labeling pipelines introduce systematic family-label noise. Second, the lack of temporally grounded expert verification limits reliable evaluation across extended time spans. As a result, it remains unclear how well malware family classifiers perform when labels are fully trustworthy, and how much of reported downstream performance is driven by the supervision source rather than the model itself.

\newcommand{\cmark}{$\checkmark$}
\newcommand{\xmark}{$\times$}
 
\setlength{\textfloatsep}{8pt}
\begin{table}[ht]
\centering
\caption{Comparison of representative Android malware datasets.}
\label{tab:malware_datasets}
\resizebox{\textwidth}{!}{%
\begin{tabular}{@{} l c r c c c c c @{}}
\toprule
\textbf{Dataset} 
  & \textbf{Years} 
  & \textbf{Samples} 
  & \textbf{Families} 
  & \textbf{Label Source} 
  & \textbf{APK}
  & \textbf{Additional Artifacts}
  & \textbf{Expert-Verified} \\
\midrule
MalGenome~\cite{zhou2012dissecting}
  & 2010--2011 & 1,260   & 49
  & AVs, bulletins, blogs
  & \cmark & ---
  & \xmark \\
Drebin~\cite{DBLP:conf/ndss/ArpSHGR14,spreitzenbarth2013mobile}
  & 2010--2012 & 5,560   & 179
  & Single AV (Kaspersky)
  & \cmark & ---
  & \xmark \\
AMD~\cite{wei2017deep}
  & 2010--2016 & 24,553  & 71
  & VT (AVClass-like)
  & \cmark & ---
  & \xmark \\
CIC-AndMal2017~\cite{lashkari2018toward}
  & 2015--2017 & 4,354   & 42
  & VT + reports/blogs
  & \cmark & NetFlow
  & \xmark \\
CCCS-CIC-AndMal-2020~\cite{rahali2020didroid}
  & 2017--2018 & 200K    & 191
  & 70\% VT consensus
  & \xmark & Static + Dynamic features$^{\dagger}$
  & \xmark \\
MalRadar~\cite{wang2022malradar}
  & 2014--2021 & 4,534   & 148
  & Reports (+ AVClass)$^{*}$
  & \cmark & ---
  & \xmark \\
MalNet~\cite{DBLP:conf/nips/FreitasDNC21}
  & 2006--2021 & 1,262,024$^{\ddagger}$ & 696
  & Engine agg. (Euphony)
  & \xmark & FCG
  & \xmark \\
BCG~\cite{hossain2025better}
  & 2017--2025 & 10,057$^{\S}$ & 126
  & VT + AVClass
  & \cmark & FCG + APK features
  & \xmark \\
\midrule
\textbf{AndroTruth (Ours)}
  & \textbf{2016--2025} & \textbf{8,172} & \textbf{187}
  & \textbf{Expert reports only}
  & $\bm{\checkmark}$ & \textbf{Static + Dynamic reports}
  & $\bm{\checkmark}$ \\
\bottomrule
\end{tabular}%
}%
\vspace{3pt}
\noindent\parbox{\textwidth}{%
  \scriptsize
  $^{*}$ MalRadar uses security reports as the primary label source but supplements with AVClass aggregation; not fully expert-verified.\\
  $^{\dagger}$ Features extracted from emulated execution; raw APK files are not released.\\
  $^{\ddagger}$ MalNet includes 79,749 benign function call graphs.\\
  $^{\S}$ BCG includes 5,986 benign APKs.
}
\end{table}

To address this gap, we introduce \textbf{AndroTruth}, an Android malware family benchmark whose labels are derived exclusively from expert technical analysis reports rather than automated voting schemes. The dataset spans 2016 to 2025 and contains 8,172 samples from 187 malware families. As shown in Table~\ref{tab:malware_datasets}, AndroTruth distinguishes itself from prior benchmarks by jointly combining released APK files, an expert-only label source, and released static/dynamic analysis artifacts. In addition, AndroTruth provides explicit temporal metadata for time-aware analysis and family-level dynamic analysis evidence for behavioral cross-checking, as described in Section~\ref{sec:method}. Our goal is not to build another larger VirusTotal-labeled corpus or a new malware classifier, but to provide a provenance-centered benchmark for studying label reliability, temporal robustness, and multimodal malware analysis.

Using AndroTruth, we conduct both label-level and downstream evaluations. Throughout the paper, GT denotes the expert-report-derived reference label selected by our documented reconciliation procedure; it is an operational experimental reference, not an assertion that expert attribution is always unique or infallible. At the label level, we identify \textit{misleading consensus}, in which automated tools agree with each other yet jointly conflict with this reference. At the downstream level, we compare two representative classification frameworks under reference-label and VirusTotal-derived supervision. In summary, this work makes the following contributions.

\begin{itemize}
    \item We construct and release \textbf{AndroTruth}\footnote{\url{https://github.com/bhpiamnothing/AndroTruth-Dataset}.}, which, to the best of our knowledge, is the largest publicly described Android malware family benchmark built entirely from expert-traceable technical reports. Prior benchmarks do not jointly provide released APK files, expert-traceable family labels, temporal metadata, and dynamic analysis artifacts, making AndroTruth a high-confidence testbed for studying family-label reliability and multimodal feature construction.

    \item Using AndroTruth, we identify and quantify \textbf{misleading consensus}, a systematic failure mode in which automated labeling tools produce the same family label for a sample, yet that label disagrees with expert ground truth. This provides direct evidence that consensus-based family labeling can propagate structured errors rather than merely producing isolated naming inconsistencies (Section~\ref{subsec:tool_label_evaluation}).

    \item We conduct downstream evaluations under expert-verified and real-world AV-derived supervision, and derive practical guidance for third-party dataset construction. In particular, we show when ClarAVy confidence-aware filtering can preserve family grouping quality and when expert-verified labels remain necessary for exact family naming.
\end{itemize}

\section{Related work}
\label{sec:related}

\subsection{Representative Android malware datasets}
Early datasets such as MalGenome~\cite{zhou2012dissecting} and Drebin~\cite{DBLP:conf/ndss/ArpSHGR14} established foundational corpora for Android malware research, but both are now dated and limited in scale. AMD~\cite{wei2017deep} expanded coverage using a VT-based aggregation strategy similar to AVClass. CIC-AndMal2017~\cite{lashkari2018toward} provides APK files and NetFlow traces, with family labels drawn from  VirusTotal, security blogs, and research reports. MalRadar~\cite{wang2022malradar} moves closer to expert knowledge by using security reports as the primary label source, but it still supplements them with AVClass aggregation.

Recent datasets increasingly emphasize scale or representation-learning convenience through derived artifacts. CCCS-CIC-AndMal-2020~\cite{rahali2020didroid} releases static and dynamic features extracted from emulated execution, but not raw APK files, under a 70\% VT-consensus labeling strategy. MalNet~\cite{DBLP:conf/nips/FreitasDNC21} focuses on graph representation learning by releasing a large collection of function call graphs without raw APK files. BCG~\cite{hossain2025better} likewise emphasizes graph-based learning through function call graphs, but unlike MalNet, it also provides APK files and APK-level features. Overall, existing datasets still derive family annotations wholly or partly from antivirus outputs or engine aggregation, and therefore remain vulnerable to naming inconsistency, temporal label drift, and systematic family-label noise.

\subsection{Label reliability and Automatic labeling tools}
The reliability of malware family labels depends critically on how they are assigned~\cite{guo2024multimodal,gao2023obfuscation}. Because antivirus vendors often disagree on family names for the same sample, researchers commonly use automatic labeling tools to normalize vendor labels. AVClass~\cite{sebastian2016avclass,sebastian2020avclass2}, the de facto standard, performs token filtering, generic-token removal, and alias detection to derive family names from antivirus outputs. ClarAVy~\cite{joyce2025claravy} improves alias resolution through tag-based classification and also provides per-sample confidence scores for confidence-aware filtering. However, both tools still aggregate engine outputs rather than consult verified ground truth. When several engines share the same misclassification, these tools may propagate rather than correct the error. This limitation is central to our study, and we quantify it directly in Section~\ref{subsec:tool_label_evaluation}.

\subsection{Positioning of AndroTruth}
Existing Android malware benchmarks generally optimize for scale, convenience, or released auxiliary representations, but still rely on AV-derived family labels or provide limited support for reliability-oriented family-level evaluation. AndroTruth is positioned as a benchmark specifically designed for label-reliability research in Android malware family analysis. It combines released APK files, an expert-only family label source, explicit temporal metadata, and released static and dynamic analysis artifacts, enabling the study of label reliability, temporal robustness, and multimodal malware analysis within a single benchmark.

\section{AndroTruth}\label{sec:method}

AndroTruth is constructed through a verifiable four-step pipeline consisting of report selection, information extraction, sample collection, and data enrichment. The guiding principle is to favor traceability and conservative uncertainty handling over maximal scale, so samples without auditable evidence or confidently reconcilable family assignments are excluded rather than force merged. Students and researchers with cybersecurity experience, including the co-authors, collected technical materials from established security vendors and independent analysts.

\subsection{Dataset Construction}\label{dataset_construction}

During report selection, each report must satisfy three requirements. (1) It explicitly states a family name for one or more Android samples. (2) It provides a sample-level Indicator of Compromise (IoC), such as a SHA-256 or MD5 hash, that uniquely identifies each candidate artifact. (3) It was published between January 2016 and the dataset construction deadline by an attributable security vendor or independent analyst and contains sufficient technical context for the curation team to audit the attribution.

Because the same sample may appear in multiple reports under different family names, we perform an explicit label-reconciliation procedure before finalizing annotations. Candidate matches are compared using IoCs, known family aliases, report dates, and documented malicious behaviors. When multiple reports are consistent after reconciliation, we retain the earliest report to preserve temporal accuracy. When a conflict cannot be resolved with sufficient confidence, we discard the sample rather than force a harmonized label. A representative case is the sample with MD5 \texttt{c907d74ace51cec7cb53b0c8720063e1}, which is labeled \textit{Youzicheng} in a Kaspersky report but \textit{Cookiethief} in an IBM report; the two names have no documented alias relationship and the reports describe substantially different behavioral foci, so we were unable to determine which labeling correctly reflects the sample's primary behavior. This case demonstrates that expert reports are traceable but fallible evidence. Our conservative policy reduces forced assignments, but cannot eliminate residual expert bias or taxonomy ambiguity. 

In the information extraction step, we adopt a human-in-the-loop procedure to verify key metadata from the selected reports. Although automated scripts are effective for retrieving candidate reports and extracting obvious fields, final family-label verification often requires cross-report semantic reconciliation that is difficult to perform reliably in a fully automated way. Reports may use inconsistent aliases, omit sufficient IoCs, or describe derivative variants under different family names. Under such conditions, automatic merging risks force-aligning distinct families or overlooking unresolved conflicts. We therefore verify and extract sample hashes (prioritizing SHA-256), family labels, malicious behaviors (e.g., ``SMS Theft'' or ``RAT''), report dates, and other structured metadata for each selected report, while also summarizing key unstructured behavioral descriptions to support downstream analysis.

In the sample collection step, we use automated scripts to retrieve samples from the Koodous repository~\cite{koodous_platform} based on the extracted hashes, and we retain only those entries for which the APK artifact can be successfully obtained. Table~\ref{tab:my_sources_summary_top20} reports the Top-20 data sources in the final dataset, including McAfee, Kaspersky, and Trend Micro, together with the number of samples contributed by each source.

\begin{table}[ht]
\centering
\caption{Distribution of malware samples and families across the Top-20 dataset sources.}
\label{tab:my_sources_summary_top20}
\begin{tabular}{lrr @{\hspace{0.5cm}} | @{\hspace{0.5cm}} lrr}
\toprule
\textbf{Source} & \textbf{Samples} & \textbf{Families} & \textbf{Source} & \textbf{Samples} & \textbf{Families} \\
\midrule
Zimperium          & 2,370 & 15 & Cybereason   & 125 & 1 \\
Lookout            & 1,840 & 14 & Zscaler      & 124 & 5 \\
Trendmicro         & 1,319 & 32 & Fortinet     & 119 & 10 \\
Kaspersky          & 380   & 36 & ThreatFabric & 116 & 23 \\
Checkpoint         & 301   & 14 & McAfee       & 99  & 10 \\
Palo Alto Networks & 228   & 8  & ESET         & 85  & 11 \\
Avast              & 156   & 2  & Clearfy      & 83  & 8  \\
IBM                & 144   & 7  & Bitdefender  & 58  & 3  \\
Cyble              & 139   & 29 & Sophos       & 55  & 6  \\
Symantec           & 134   & 1  & Group-IB     & 54  & 1  \\
\bottomrule
\end{tabular}
\end{table}

Finally, in the data enrichment step, we query the VirusTotal API~\cite{10.1145/3711896.3737431,jin2024sharing} to obtain the first submission date of each sample and align it with the corresponding report date. This alignment links each sample's initial appearance, subsequent vendor detection, and eventual public disclosure, thereby enabling later analysis of detection latency and malware evolution.

Across the four pipeline steps, we collected a total of 278 security reports and blog posts, from which we extracted 8{,}421 unique sample hashes after cross-report de-duplication. Among these, 245 hashes could not be retrieved from Koodous and were therefore discarded and not included in the dataset. In addition, 4 samples were excluded during label reconciliation because their cross-report family assignments could not be unambiguously resolved (as illustrated by the \textit{Youzicheng/Cookiethief} case discussed above). After these filtering steps, the final AndroTruth release contains 8{,}172 samples from 187 malware families, drawn from 276 security reports and blog posts published by 42 vendors and individual analysts between February 2016 and November 2025.

\subsection{Families and Samples Distribution}
Table~\ref{tab:my_sources_summary_top20} also reveals substantial variation in source-level sample diversity. Zimperium and Lookout contribute large numbers of samples from relatively few families, whereas Kaspersky, ThreatFabric, and Cyble contribute fewer samples but cover a broader set of families. Thus, the sources provide complementary forms of coverage, with some offering greater within-family depth and others increasing cross-family breadth.
 
\subsubsection{Temporal Distribution.}
\begin{figure}[ht]
  \centering
  \includegraphics[width=0.8\columnwidth]{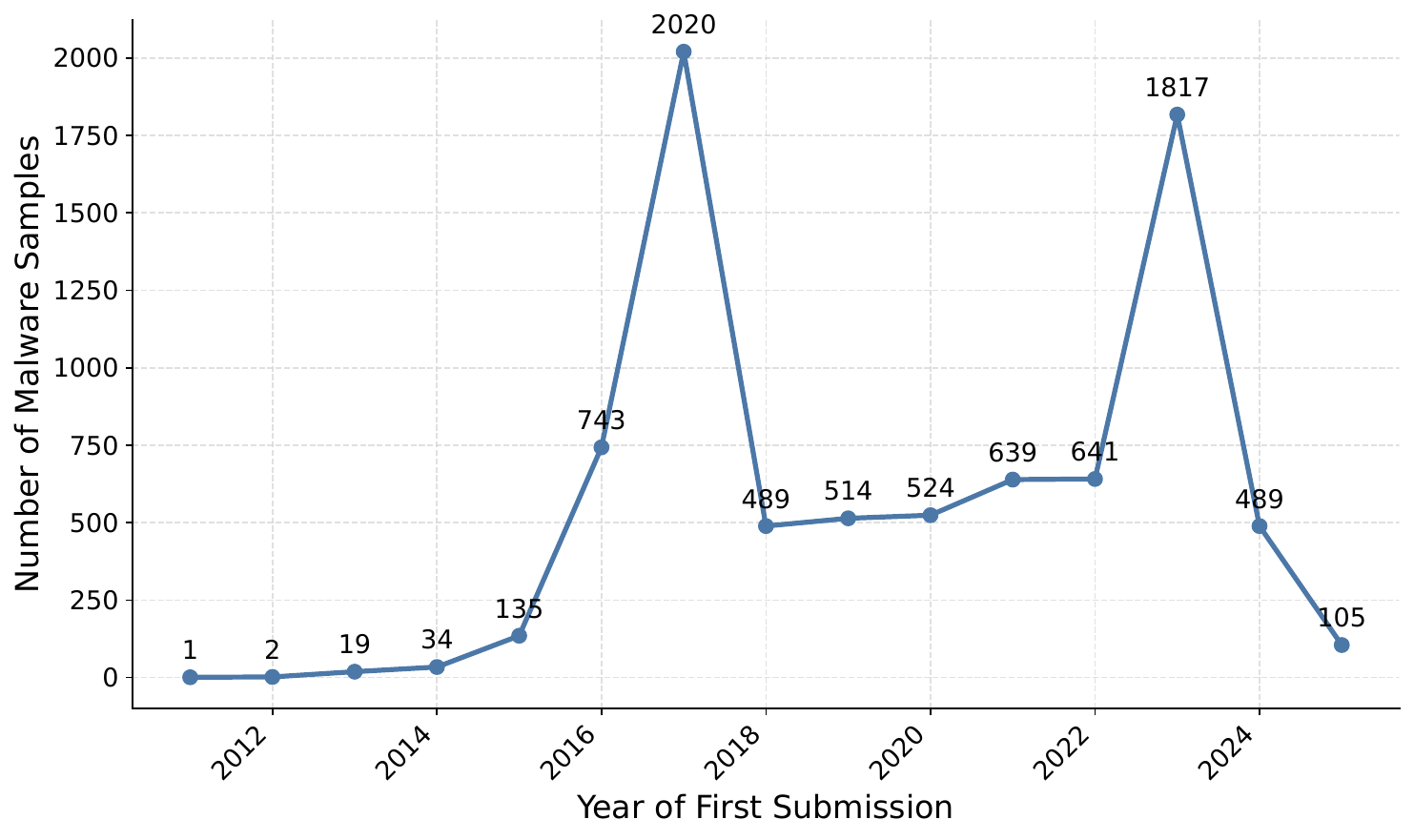}
  \caption{Temporal distribution of malware samples based on first submission dates by VirusTotal.}
  \label{fig:report_timeline}
\end{figure}

We characterize the temporal coverage of AndroTruth using the ``first submission date'' of each sample reported by VirusTotal. Figure~\ref{fig:report_timeline} shows a bimodal distribution, with a first peak in 2017 (2,020 samples), a comparatively stable period from 2018 to 2022, and a second peak in 2023 (1,817 samples), driven largely by the \textit{Godfather} family. Rather than reflecting a single short-lived campaign, the dataset spans nearly a decade and includes many recent samples, making it suitable for studying temporal drift and long-term changes in Android malware activity~\cite{11025877}.
 
\begin{figure}[ht]
  \centering
  \includegraphics[width=0.8\columnwidth]{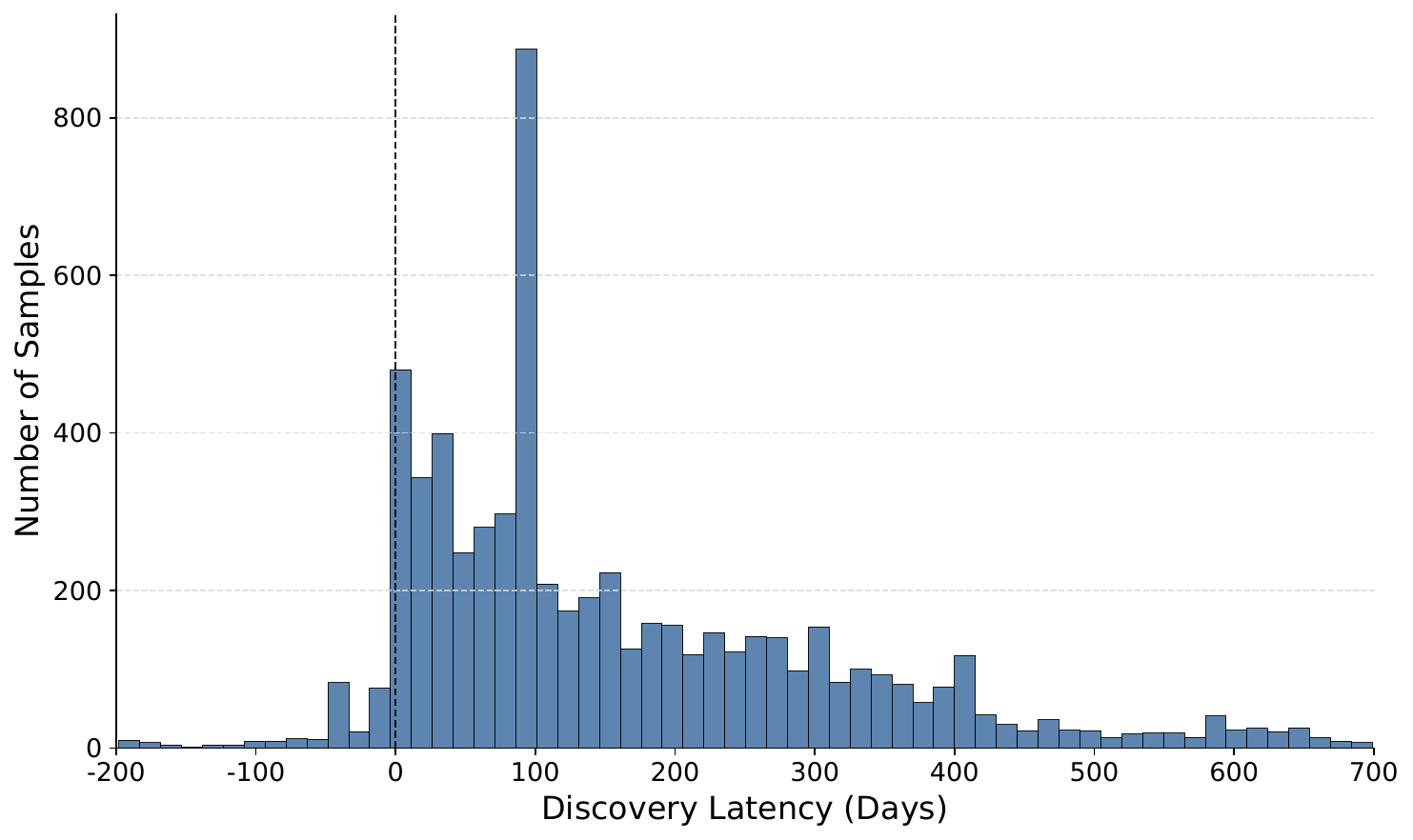}
  \caption{Time between first VirusTotal submission and public expert report disclosure.}
  \label{fig:discovery_latency}
\end{figure}

The temporal span also allows us to characterize reporting latency, defined as the interval between a sample's first VirusTotal submission and its public disclosure in an expert report. As shown in Figure~\ref{fig:discovery_latency}, 10.74\% of samples appear in public reports within 30 days of their initial submission, 28.63\% within 90 days, and 28.34\% more than one year later. These intervals measure public-report timing relative to the VirusTotal timestamp; they do not by themselves establish that a sample was dormant, highly stealthy, or undetected. Delayed disclosure, internal vendor telemetry, campaign-level reporting, and incomplete alignment between collection systems are plausible alternatives. Approximately 6.03\% of samples exhibit negative latency, which is consistent with internal collection or later VirusTotal upload, but is not sufficient evidence that every such sample was a zero-day threat.

\subsubsection{Family Size Distribution.}
\begin{figure}[ht]
  \centering
  \includegraphics[width=0.8\columnwidth]{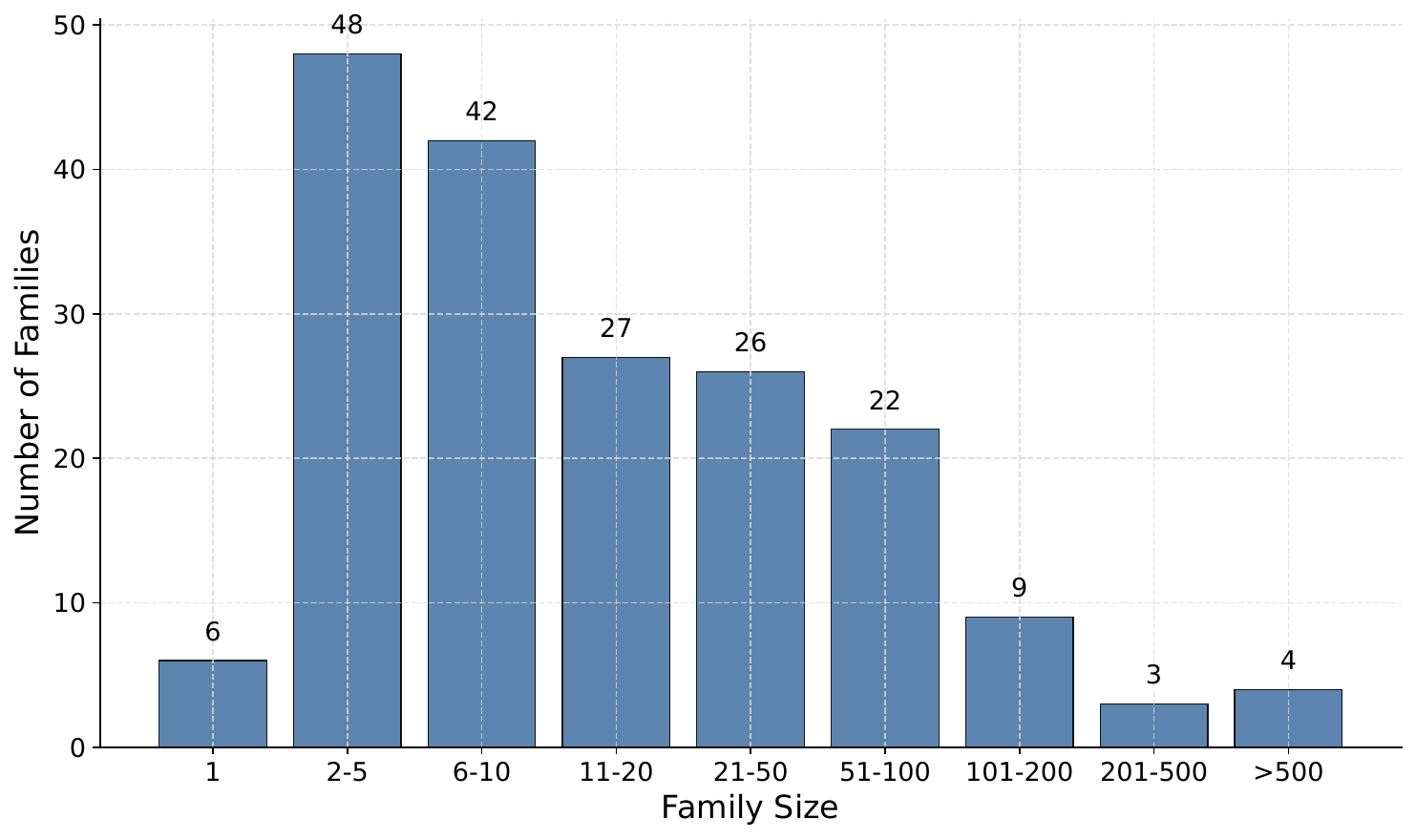}
  \caption{Distribution of malware family sizes.}
  \label{fig:family_size_distribution}
\end{figure}

Figure~\ref{fig:family_size_distribution} shows a pronounced long-tail family-size distribution. The top ten families account for 52.48\% of all samples, while the largest family, \textit{Godfather}, alone contributes 1,173 samples (14.35\%). At the other extreme, 91 families contain fewer than 10 samples. This distribution accurately reflects the imbalance observed in real-world malware ecosystems~\cite{bai2020unsuccessful} and creates a realistic setting for evaluating learning methods under class imbalance.
 
\subsubsection{Malicious Behavior Categories.}
To summarize threat diversity, we categorize AndroTruth according to Kaspersky's official classification standards\footnote{\url{https://encyclopedia.kaspersky.com/knowledge/classification/}.}. This taxonomy provides a practical behavioral view of the corpus and is useful for explainable Android malware analysis~\cite{9978981,10.1145/3677374}.
 
Table~\ref{tab:category_detailed_generated} presents the distribution across different categories. \textit{Trojan-Banker} ranks first with a share of 39.59\%, represented mainly by the \textit{Godfather} family, which targets financial and banking applications. This reflects the prevalence of financially motivated attacks. \textit{Trojan-Spy} follows at 28.77\% and primarily includes spyware that steals private data such as SMS messages, contacts, and location information. In addition, \textit{Backdoor} (6.56\%), generic \textit{Trojan} (5.10\%), and \textit{Trojan-Dropper} (4.52\%) constitute substantial portions of the dataset. Smaller categories, including \textit{Trojan-Clicker}, \textit{Adware}, and \textit{Ransomware}, occupy lower proportions but increase the behavioral diversity of the dataset. Threats targeting financial assets (\textit{Trojan-Banker}) and private data (\textit{Trojan-Spy}) together account for nearly 70\% of all samples, indicating that current Android malware predominantly focuses on high-value targets.
 
\begin{table}[ht]
\centering
\caption{Distribution of malware categories.}
\label{tab:category_detailed_generated}
\small
\begin{tabular}{@{} l r r @{\hspace{14pt}} l r r @{}}
\toprule
\textbf{Category} & \textbf{Count} & \textbf{\%} & \textbf{Category} & \textbf{Count} & \textbf{\%} \\
\midrule
Trojan-Banker            & 3{,}235 & 39.59 & Trojan-Spy;Trojan-Downloader & 99  & 1.21 \\
Trojan-Spy               & 2{,}351 & 28.77 & Trojan-Ransom                & 97  & 1.19 \\
Backdoor                 & 536     & 6.56  & Trojan-Banker;Trojan-Ransom  & 72  & 0.88 \\
Trojan                   & 417     & 5.10  & Backdoor;Trojan-Spy          & 66  & 0.81 \\
Trojan-Dropper           & 369     & 4.52  & Trojan-Downloader            & 41  & 0.50 \\
Trojan-Clicker           & 293     & 3.59  & Trojan-SMS;Trojan-Clicker    & 20  & 0.24 \\
Trojan-Spy;Backdoor      & 154     & 1.88  & RiskTool                     & 12  & 0.15 \\
Trojan-SMS               & 153     & 1.87  & Trojan-FakeAV                & 3   & 0.04 \\
Adware                   & 131     & 1.60  & Trojan-PSW                   & 2   & 0.02 \\
Adware;Trojan-Clicker    & 119     & 1.46  & Trojan-Proxy                 & 2   & 0.02 \\
\bottomrule
\end{tabular}
\end{table}


\subsection{Analysis of VirusTotal Scan Results}

We describe cross-family variation in the number of antivirus engines that flagged each sample in the available VirusTotal scan result. Table~\ref{tab:detection_extremes} reports statistics for the ten families with the highest and lowest mean detection counts, revealing substantial variation in observed scan coverage.

Highly recognized families such as \textit{Cookiethief} are detected by more than 42 engines even in their hardest cases, whereas evasive families such as \textit{RelayNFC} and \textit{Gigabud} receive far fewer detections, with some \textit{Gigabud} samples receiving none at all. This polarization is valuable for researchers in several ways. First, detection count serves as a proxy for evasion difficulty, enabling researchers to stratify evaluation by threat sophistication rather than treating all families as equally challenging. Second, families with persistently low detection rates highlight blind spots in current antivirus engines, guiding future work on evasion-resistant detection methods. Third, including both extremes ensures that AndroTruth supports realistic benchmarking scenarios in which classifiers must handle not only well-known threats but also stealthy samples that evade the majority of existing engines.

\begin{table}[ht]
\centering
\caption{Top 10 most-detected and least-detected malware families by antivirus engines.}
\label{tab:detection_extremes}
\small
\begin{tabular}{@{} l r r r r @{\hspace{12pt}} l r r r r @{}}
\toprule
\multicolumn{5}{c}{\textbf{Most Detected}} & \multicolumn{5}{c}{\textbf{Least Detected}} \\
\cmidrule(r){1-5} \cmidrule(l){6-10}
\textbf{Family} & \textbf{Mean} & \textbf{Med.} & \textbf{Max} & \textbf{Min}
& \textbf{Family} & \textbf{Mean} & \textbf{Med.} & \textbf{Max} & \textbf{Min} \\
\midrule
Cookiethief  & 43.33 & 44.0 & 44 & 42 & RelayNFC    & 13.20 & 13.0 & 14 &  12 \\
Hiddenminer  & 42.58 & 42.5 & 47 & 40 & Enchant     & 16.00 & 16.0 & 19 &  13 \\
Funkybot     & 39.50 & 39.5 & 42 & 37 & Necro       & 20.01 & 22.0 & 33 &   3 \\
Spydealer    & 38.00 & 39.5 & 41 & 32 & Fakebank    & 20.18 & 20.0 & 23 &  16 \\
BugDrop      & 38.00 & 38.0 & 38 & 38 & PixPirate   & 20.39 & 22.0 & 38 &   7 \\
Nexus        & 38.00 & 38.0 & 38 & 38 & Gigabud     & 20.64 & 24.0 & 44 &   0 \\
Zniu         & 37.90 & 41.0 & 42 & 23 & Antidot     & 21.33 & 22.0 & 22 &  20 \\
Sova         & 37.00 & 36.5 & 40 & 35 & Cherryblos  & 21.50 & 20.5 & 26 &  18 \\
Kevdroid     & 36.67 & 37.0 & 39 & 34 & AppLite     & 21.52 & 22.0 & 28 &  13 \\
Bianlian     & 36.00 & 36.0 & 36 & 36 & Facestealer & 21.64 & 21.5 & 27 &  16 \\
\bottomrule
\end{tabular}
\end{table}

\subsection{Evaluation of Labels Produced by Automated Tools}
\label{subsec:tool_label_evaluation}

Automated family labeling tools such as AVClass2~\cite{sebastian2020avclass2} and ClarAVy~\cite{joyce2025claravy} aggregate VirusTotal scan results and are widely used in Android malware research. However, this engine-consensus paradigm introduces substantial label noise. We therefore evaluate tool-derived family labels against the expert-verified labels of AndroTruth, which serve as ground truth (GT). As described in Section~\ref{dataset_construction}, every GT label is traceable to documented forensic evidence from expert technical analysis reports. This traceability makes expert labels more suitable as ground truth than tool-derived labels, which lack explicit forensic context or auditable evidence trails. 

To avoid conflating incorrect family assignments with absent family assignments, we distinguish \emph{explicit family labels} from \emph{tool-unlabeled outcomes}. For AVClass2 and ClarAVy, outputs such as \texttt{benign} are treated as \emph{no malicious detection}, whereas \texttt{SINGLETON} is treated as an unresolved malware case without a reliable family assignment. For Kaspersky, \texttt{null} is likewise treated as \emph{no malicious detection}, while \texttt{N/A} is treated as \emph{missing scan}. None of these outcomes is counted as an ordinary family prediction. Table~\ref{tab:label_reliability} reports the resulting label-audit statistics on the aligned set of 8,172 GT malware samples. We highlight several findings from this analysis. 

\textbf{High explicit family coverage does not imply high label reliability}. Kaspersky and ClarAVy both provide explicit family labels for more than 95\% of the samples, yet Kaspersky conflicts with GT on 55.20\% of its labeled subset, substantially worse than AVClass2 (42.51\%) and ClarAVy (37.11\%). Among the automated tools, ClarAVy achieves the best overall trade-off between coverage and correctness, while AVClass2 produces a noticeably larger proportion of \texttt{SINGLETON} outputs (8.84\%), indicating weaker family resolution ability.
\begin{table*}[ht]
\centering
\caption{Label reliability analysis before any classifier.}
\label{tab:label_reliability}
\setlength{\tabcolsep}{6pt}
\resizebox{\textwidth}{!}{%
\begin{tabular}{lcccccc}
\toprule
\textbf{Source} & \textbf{Explicit family} & \textbf{Singleton} & \textbf{No mal. det.} & \textbf{Missing scan} & \textbf{Match w/ GT} & \textbf{Conflict w/ GT} \\
\midrule
Kaspersky & 7,814 (95.62\%) & 0 (0.00\%) & 250 (3.06\%) & 108 (1.32\%) & 3,501 / 7,814 (44.80\%) & 4,313 / 7,814 (55.20\%) \\
AVClass2  & 7,413 (90.71\%) & 722 (8.84\%) & 37 (0.45\%) & 0 (0.00\%) & 4,262 / 7,413 (57.49\%) & 3,151 / 7,413 (42.51\%) \\
ClarAVy   & 7,795 (95.39\%) & 340 (4.16\%) & 37 (0.45\%) & 0 (0.00\%) & 4,902 / 7,795 (62.89\%) & 2,893 / 7,795 (37.11\%) \\
\bottomrule
\end{tabular}
}
\end{table*}

Beyond coverage, \textbf{the tools also exhibit distinct failure patterns}. Kaspersky produces both \texttt{no malicious detection} and \texttt{missing scan} cases, and when it does emit a family label it often favors coarse-grained vendor-style names. By contrast, AVClass2 and ClarAVy rarely miss scans in the aligned subset, but they more often fail through unresolved \texttt{SINGLETON} outputs or incorrect explicit family assignments.

Most importantly, \textbf{agreement between tools is not evidence of correctness}. On the 7,243 samples where both AVClass2 and ClarAVy output explicit family labels, only 55.71\% are correct under both tools, while 33.19\% are wrong under both. Most notably, 1,838 samples (25.38\%) exhibit \emph{misleading consensus}. We define misleading consensus as the case where two or more automated tools assign the same family label to a sample, yet that label disagrees with expert ground truth. This result shows that consensus-based automated labeling pipelines can propagate structured family label errors rather than merely producing isolated naming inconsistencies.

Detailed misclassification analysis reveals three recurring failure modes across the evaluated label sources (Table~\ref{tab:tool_errors}). All three sources persistently confuse grayware and adware-related families. For instance, \textit{Ghostclicker} is mislabeled as \textit{Hiddad} by AVClass2 (209 cases), ClarAVy (201), and Kaspersky (124), while \textit{Judy} is similarly confused with \textit{Ewind} across all three tools, indicating that such overlaps are intrinsic to scan-based labeling rather than specific to any single tool. Families sharing code lineage or similar runtime behavior are also frequently conflated. \textit{AppLite} is mislabeled as \textit{Hqwar} by all three sources (152--159 cases each), and the banking trojan \textit{Godfather} is collapsed into its progenitor \textit{Cerberus} by ClarAVy (145 cases) and AVClass2 (75 cases), directly distorting lineage analysis and threat attribution. Beyond these shared errors, Kaspersky exhibits a distinctive failure mode, frequently mapping distinct GT
families to the broad vendor-style label \textit{Agent}, including \textit{CarbonSteal} (270 cases), \textit{Henbox} (150), \textit{Godfather} (138), and \textit{GoldenEagle} (89). Unlike the inter-family confusions observed in AVClass2 and ClarAVy, which typically involve behaviorally similar families, this pattern collapses families with substantially different malicious behaviors into a single coarse-grained label. Because AVClass2 and ClarAVy derive their outputs by aggregating scan results from multiple engines including Kaspersky, such vendor-style labels may propagate into their labeling pipelines and contribute to downstream errors, particularly when other engines do not provide sufficiently specific family names to override the coarse assignment.

\begin{table*}[ht]
\centering
\caption{Top 10 misclassifications per label source (maximum error per family).}
\label{tab:tool_errors}
\small
\begin{tabular}{@{} l l r @{\hspace{10pt}} l l r @{\hspace{10pt}} l l r @{}}
\toprule
\multicolumn{3}{c}{\textbf{AVClass2}} & \multicolumn{3}{c}{\textbf{ClarAVy}} & \multicolumn{3}{c}{\textbf{Kaspersky}} \\
\cmidrule(r){1-3} \cmidrule(lr){4-6} \cmidrule(l){7-9}
\textbf{GT} & \textbf{Pred.} & \textbf{Err.}
& \textbf{GT} & \textbf{Pred.} & \textbf{Err.}
& \textbf{GT} & \textbf{Pred.} & \textbf{Err.} \\
\midrule
CarbonSteal  & Mlasdl   & 224 & CarbonSteal  & Zitmo    & 379 & CarbonSteal  & Agent    & 270 \\
Ghostclicker & Hiddad   & 209 & Ghostclicker & Hiddad   & 201 & AppLite      & Hqwar    & 152 \\
AppLite      & Hqwar    & 156 & AppLite      & Hqwar    & 159 & Henbox       & Agent    & 150 \\
UltimaSMS    & Vuad     & 130 & Godfather    & Cerberus & 145 & UltimaSMS    & Vuad     & 144 \\
Ghostctrl    & Gopnok   & 108 & Judy         & Ewind    & 112 & Godfather    & Agent    & 138 \\
Judy         & Ewind    &  92 & Ghostctrl    & Gopnok   & 108 & Ghostclicker & Hiddad   & 124 \\
SilkBean     & Fakeapp  &  82 & PixPirate    & Banbra   &  77 & Ghostctrl    & Gopnok   & 107 \\
Godfather    & Cerberus &  75 & Bankbot      & Asacub   &  62 & Judy         & Ewind    & 103 \\
PixPirate    & Banbra   &  71 & Henbox       & Hidden   &  50 & GoldenEagle  & Agent    &  89 \\
Geost        & Hqwar    &  49 & Geost        & Hqwar    &  48 & Copybara     & Basbanke &  83 \\
\bottomrule
\end{tabular}
\end{table*}

\subsection{Analysis of Static and Dynamic Features}\label{subsec:static_dynamic}

AndroTruth also releases static and dynamic analysis artifacts that make the dataset usable beyond fixed label files. These artifacts provide cross-modal evidence for family analysis and help explain why some family distinctions are difficult for scan-based tools.

\smallskip
\noindent\textbf{Static Features.}
We extract Drebin-style features from each APK using Androguard across ten categories: requested permissions, used permissions, activities, services, broadcast receivers, content providers, intent filters, restricted APIs, suspicious APIs, and embedded URLs, with permission-to-API mappings supporting up to Android API level~36. Of the 8{,}172 samples in AndroTruth, Androguard successfully parsed 8{,}145 (99.67\%) spanning 186 families; the remaining 27 samples (belonging to one small family) could not be fully decompiled due to APK packing, heavy obfuscation, or structural anomalies, and are excluded from static-feature analysis. On the retained set, each sample contains an average of 99.1 features (median:~79). The most frequent permissions include \texttt{INTERNET} (97\%), \texttt{WAKE\_LOCK} (77\%), \texttt{ACCESS\_NETWORK\_STATE} (77\%), and \texttt{READ\_PHONE\_STATE} (71\%), reflecting the network-centric behavior of Android malware at scale.

\begin{figure}[ht]
  \centering
  \includegraphics[width=\columnwidth]{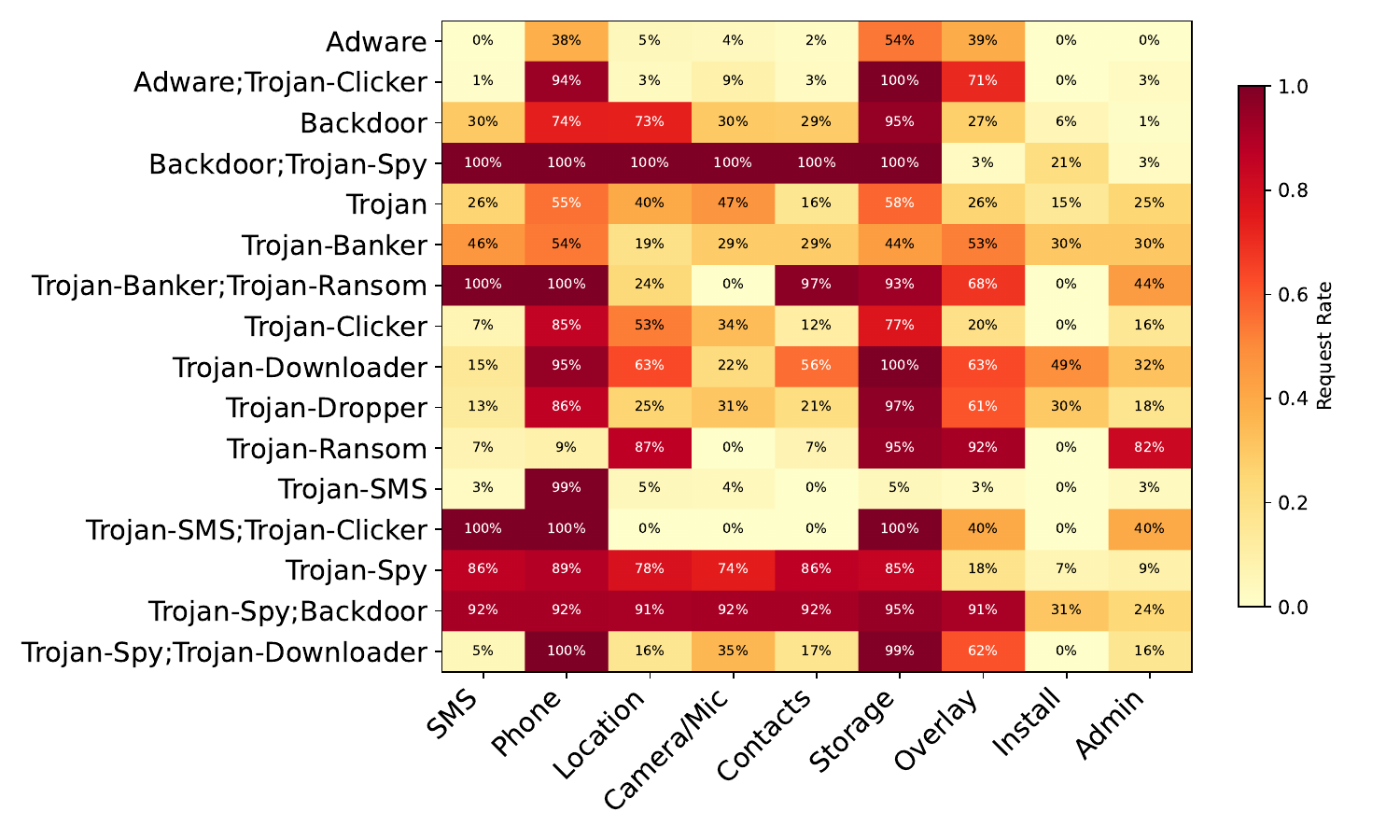}
  \caption{Sensitive permission request rate by malware category.}
  \label{fig:perm_heatmap}
\end{figure}

Figure~\ref{fig:perm_heatmap} shows clear category-specific signatures. The \textit{Trojan-Spy} category requests SMS (86\%), Phone (89\%), Location (78\%), and Contacts (86\%) permissions at high rates. The \textit{Trojan-Banker} category shows a characteristic Overlay rate of 53\%, and the \textit{Trojan-Ransom} category exhibits the highest Admin (82\%) and Overlay (92\%) rates. In contrast, the \textit{Adware} category requests few sensitive permissions. These patterns provide static evidence that broadly corroborates the expert-assigned labels.

\smallskip
\noindent\textbf{Dynamic Analysis.}
A total of 3{,}893 samples (47.64\% of the 8{,}172 samples) include structured dynamic analysis reports retrieved from the Koodous sandbox, recording runtime events across eleven categories. This partial coverage reflects rate-limit and per-account query quotas imposed by the Koodous API at the time of data collection, rather than any intrinsic selection criterion on the samples themselves; the retained subset is therefore best viewed as a random-access slice of the full corpus. Table~\ref{tab:dynamic_prevalence} shows that file reads are ubiquitous (100\%), dynamic DEX loading occurs in 51.0\% of samples, and cryptographic API usage appears in 29.9\%. Network events are captured less frequently (DNS:~30.8\%, outbound:~5.4\%), which likely reflects sandbox connectivity limitations rather than behavioral absence.

\begin{table}[ht]
\centering
\caption{Dynamic feature prevalence (3,893 sandbox-analyzed samples).}
\label{tab:dynamic_prevalence}
\small
\setlength{\tabcolsep}{10pt}
\renewcommand{\arraystretch}{1.15}
\begin{tabular}{l r @{\hspace{18pt}} l r}
\toprule
\textbf{Event Type} & \textbf{Rate} & \textbf{Event Type} & \textbf{Rate} \\
\midrule
Files Read     & 100.0\% & DNS Queries     & 30.8\% \\
Files Written  & 76.9\%  & Crypto Usage    & 29.9\% \\
DEX Classes    & 51.0\%  & Service Start   & 18.6\% \\
Libraries      & 35.4\%  & Net Recv / Send & 11.8\% / 5.4\% \\
\bottomrule
\end{tabular}
\end{table}

\smallskip

\begin{figure}[ht]
  \centering
  \includegraphics[width=\columnwidth]{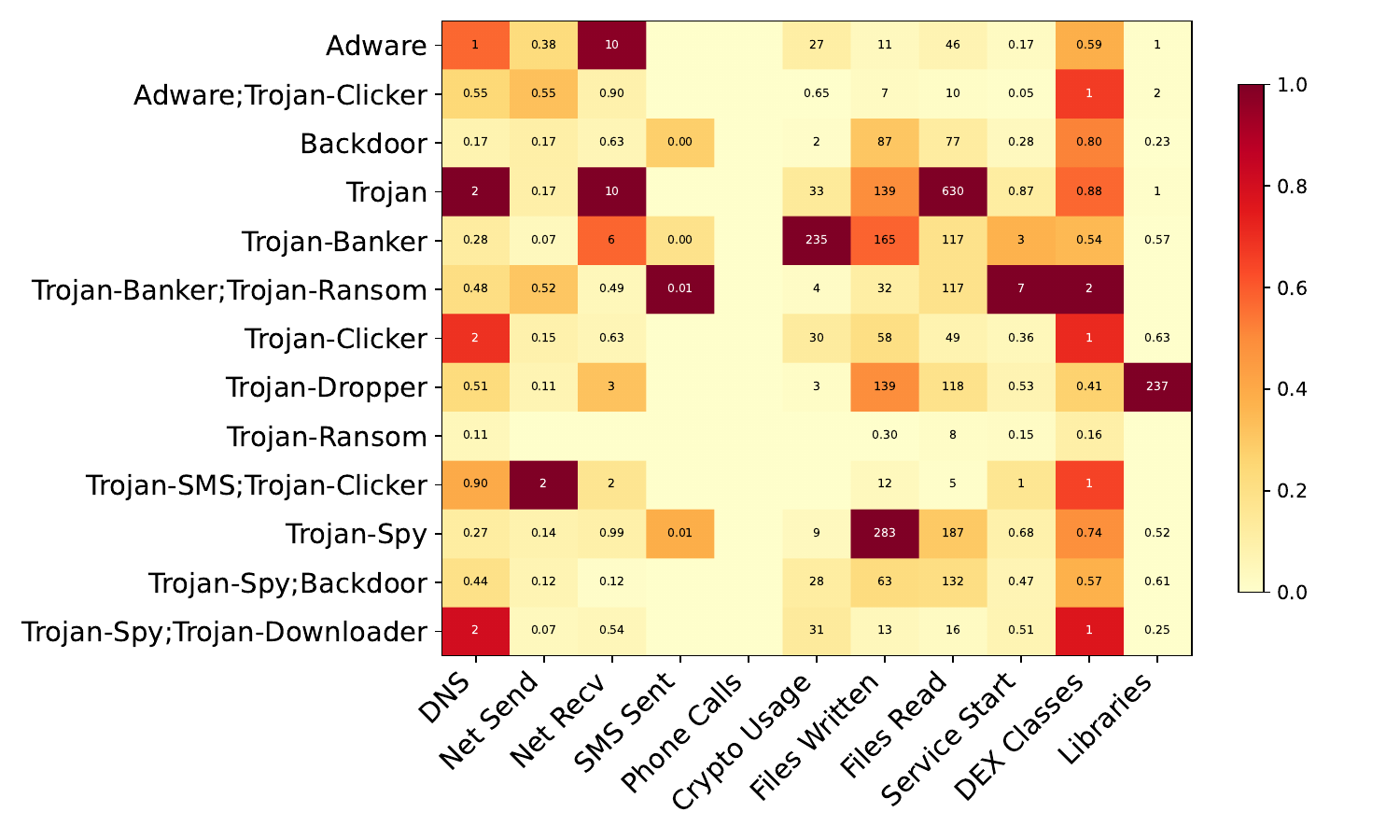}
  \caption{Normalized dynamic behavioral profile by malware category.}
  \label{fig:dynamic_heatmap}
\end{figure}

Figure~\ref{fig:dynamic_heatmap} shows category-level dynamic signatures that are broadly consistent with the static profiles above. The \textit{Trojan-Banker} category exhibits the highest cryptographic API usage (235 events per sample on average), the \textit{Trojan-Spy} category dominates in file-write volume (283 events), and the \textit{Trojan-Dropper} category shows the highest library-loading rate (237 events). Together, the static and dynamic views provide cross-modal support for the expert labels. However, these category-level profiles aggregate behavior across families, and category-level separation does not imply fine-grained family-level separation. Several of the persistently confused pairs in Table~\ref{tab:tool_errors} involve families within the same behavioral category, such as \textit{Godfather}--\textit{Cerberus} and \textit{Bankbot}--\textit{Asacub} within Trojan-Banker, whose runtime signatures in the sandbox reports are largely overlapping. This suggests that sandbox-level dynamic behavior alone often provides limited signal for fine-grained family distinctions, and that resolving such distinctions typically requires the forensic context provided by expert technical reports, precisely the evidence that underpins AndroTruth's labels.
\smallskip

\section{Experimental Design}
\label{sec:experiment}

Having established in Section~\ref{subsec:tool_label_evaluation} that automated family labels exhibit substantial annotation-level unreliability, we investigate three downstream questions.
\textbf{RQ1}: How does unreliable supervision affect family-level Android malware classification?
\textbf{RQ2}: Is the resulting degradation driven mainly by the \emph{rate} of label errors, or by their \emph{structure}?
\textbf{RQ3}: How can researchers reduce label noise when constructing third-party datasets?

To answer these questions, we use two representative open-source frameworks, AndMFC~\cite{8880840} and Meta-MAMC~\cite{10.1145/3664806}, to evaluate how different supervision conditions affect downstream classification outcomes. Our goal is not to compare the two classifiers against each other, but to measure how different supervision conditions change downstream conclusions.

\subsection{Experimental Setup}
\label{subsec:rq_setup}

\textbf{Label sources.}
We consider four label sources in the main downstream evaluation. (1) Expert-verified labels from AndroTruth as the clean reference, (2) labels from a single antivirus engine (Kaspersky), (3) AVClass2-aggregated labels, and (4) ClarAVy-aggregated labels. For noisy label sources, only samples with explicit family assignments are used for training.

\textbf{Evaluation protocol.}
We adopt a strict five-fold stratified evaluation protocol. To avoid train--test asymmetry caused by ultra-rare families, we globally exclude GT families with fewer than five samples before cross-validation. Within each fold, models are trained using one of the label sources above and are always evaluated against GT on the corresponding test split. The results in Section~\ref{subsec:downstream_label_sources} therefore reflect real-world usable supervision, in which both label quality and usable labeled coverage vary by source. Section~\ref{subsec:matched_pool_decomposition} then isolates these factors on matched shared pools.

\textbf{Metrics.}
For exact family naming, we report Accuracy (ACC), Macro Precision, Macro Recall, and Macro-F1, which require the predicted family name to match the GT name after alias reconciliation. Following FARE~\cite{DBLP:conf/ndss/Liang0LH0X21}, we additionally use Adjusted Mutual Information (AMI) to measure the overall agreement between the predicted and GT partitions after correcting for chance. Following the cluster-oriented evaluation adopted by AVClass and Euphony~\cite{sebastian2016avclass,7962391}, we also report Cluster Precision, Cluster Recall, and Cluster-F1. Unlike the exact-name metrics, these clustering-oriented metrics do not require the predicted family names to match the GT names literally. Instead, they focus on the underlying grouping structure by examining whether samples from the same GT family are kept together and whether samples placed in the same predicted cluster actually belong to the same GT family. Cluster Precision evaluates how effectively predicted clusters map to GT families, Cluster Recall evaluates how effectively GT families are recovered by predicted clusters, and Cluster-F1 is their harmonic mean. Together, these two groups of metrics distinguish naming or taxonomy differences from genuine errors in the underlying sample partition.

\subsection{Downstream Impact Under Different Label Sources}
\label{subsec:downstream_label_sources}
Table~\ref{tab:downstream_real_world} reports downstream classification performance under the four supervision sources. These results reflect real world usable supervision, where each source differs in both label quality and labeled coverage. Two observations stand out. Expert verified GT is dramatically stronger than every automated alternative, and among the noisy sources the ranking is stable across both frameworks, with ClarAVy performing best, followed by AVClass2 and then Kaspersky.

\begin{table*}[ht]
\centering
\caption{Downstream performance under different label sources.}
\label{tab:downstream_real_world}
\setlength{\tabcolsep}{6pt}
\resizebox{\textwidth}{!}{%
\begin{tabular}{llcccc}
\toprule
\textbf{Method} & \textbf{Label Source} & \textbf{ACC} & \textbf{Macro-Precision} & \textbf{Macro-Recall} & \textbf{Macro-F1} \\
\midrule
\multirow{4}{*}{AndMFC} & GT & 96.66\% $\pm$ 0.40\% & 91.73\% $\pm$ 1.39\% & 90.50\% $\pm$ 1.21\% & 90.20\% $\pm$ 1.42\% \\
 & Kaspersky & 43.52\% $\pm$ 0.74\% & 21.80\% $\pm$ 1.13\% & 21.50\% $\pm$ 1.23\% & 20.42\% $\pm$ 1.13\% \\
 & AVClass2 & 54.46\% $\pm$ 0.93\% & 33.74\% $\pm$ 1.06\% & 31.03\% $\pm$ 0.74\% & 29.90\% $\pm$ 0.77\% \\
 & ClarAVy & 59.72\% $\pm$ 0.66\% & 36.90\% $\pm$ 1.09\% & 35.85\% $\pm$ 0.53\% & 34.44\% $\pm$ 0.74\% \\
\midrule
\multirow{4}{*}{Meta-MAMC} & GT & 96.92\% $\pm$ 0.34\% & 88.51\% $\pm$ 1.41\% & 88.30\% $\pm$ 1.17\% & 87.30\% $\pm$ 1.38\% \\
 & Kaspersky & 44.34\% $\pm$ 0.69\% & 20.41\% $\pm$ 1.08\% & 20.85\% $\pm$ 0.68\% & 19.50\% $\pm$ 0.79\% \\
 & AVClass2 & 55.77\% $\pm$ 1.01\% & 33.08\% $\pm$ 1.08\% & 31.47\% $\pm$ 1.05\% & 30.09\% $\pm$ 0.96\% \\
 & ClarAVy & 60.27\% $\pm$ 0.72\% & 37.37\% $\pm$ 0.96\% & 36.72\% $\pm$ 0.85\% & 34.93\% $\pm$ 0.87\% \\
\bottomrule
\end{tabular}
}
\end{table*}

When GT supervision is replaced by real-world noisy labels, performance drops sharply. The gap is especially severe for Kaspersky-derived labels, where Macro-F1 falls to about 20\% for both probes. AVClass2 improves over the single-engine baseline, and ClarAVy is consistently the strongest noisy source, but both remain far below the clean-label reference.

Table~\ref{tab:downstream_clustering} shows the same ordering in clustering-oriented metrics. The simultaneous drop in Macro-F1, AMI, and Cluster-F1 indicates that the degradation is not merely a family-name vocabulary issue; the underlying family partition is also corrupted. Across these two static-feature-based probes, supervision quality has a larger effect than probe choice. This relative ordering is robust to several additional settings; Appendix~\ref{app:singleton_sensitivity} shows that the same conclusion holds when \texttt{SINGLETON} outputs are retained as a unified pseudo-label under a matched-sample fair protocol.

\begin{table*}[ht]
\centering
\caption{Clustering-oriented metrics under different label sources.}
\label{tab:downstream_clustering}
\setlength{\tabcolsep}{6pt}
\resizebox{\textwidth}{!}{%
\begin{tabular}{llcccc}
\toprule
\textbf{Method} & \textbf{Label Source} & \textbf{AMI} & \textbf{Cluster-Precision} & \textbf{Cluster-Recall} & \textbf{Cluster-F1} \\
\midrule
\multirow{4}{*}{AndMFC} & GT & 96.46\% $\pm$ 0.51\% & 96.71\% $\pm$ 0.39\% & 97.27\% $\pm$ 0.38\% & 96.99\% $\pm$ 0.39\% \\
 & Kaspersky & 73.26\% $\pm$ 0.81\% & 73.94\% $\pm$ 0.73\% & 78.82\% $\pm$ 0.73\% & 76.30\% $\pm$ 0.68\% \\
 & AVClass2 & 81.45\% $\pm$ 0.67\% & 85.28\% $\pm$ 0.58\% & 79.94\% $\pm$ 1.05\% & 82.52\% $\pm$ 0.75\% \\
 & ClarAVy & 85.61\% $\pm$ 0.89\% & 89.63\% $\pm$ 0.85\% & 82.99\% $\pm$ 0.68\% & 86.18\% $\pm$ 0.69\% \\
\midrule
\multirow{4}{*}{Meta-MAMC} & GT & 97.16\% $\pm$ 0.30\% & 97.04\% $\pm$ 0.32\% & 97.81\% $\pm$ 0.30\% & 97.42\% $\pm$ 0.31\% \\
 & Kaspersky & 74.67\% $\pm$ 1.31\% & 74.73\% $\pm$ 1.15\% & 80.54\% $\pm$ 0.93\% & 77.53\% $\pm$ 1.04\% \\
 & AVClass2 & 83.93\% $\pm$ 0.21\% & 86.22\% $\pm$ 0.50\% & 83.02\% $\pm$ 0.62\% & 84.59\% $\pm$ 0.28\% \\
 & ClarAVy & 87.24\% $\pm$ 0.69\% & 90.35\% $\pm$ 0.79\% & 84.33\% $\pm$ 0.59\% & 87.24\% $\pm$ 0.68\% \\
\bottomrule
\end{tabular}
}
\end{table*}

\subsection{Matched-Pool Decomposition of Noisy Supervision}
\label{subsec:matched_pool_decomposition}

The previous section establishes that noisy labels degrade classification, but does not reveal whether the damage stems from error rate or error structure (\textbf{RQ2}). The matched-pool analysis separates label coverage, conflict rate, and conflict destination. Uniform corruption replaces a selected GT label with a randomly chosen wrong family. Structured corruption uses the same overall rate but draws the replacement family from the source-specific confusion distribution estimated on the training partition. If structured corruption is more damaging at a matched rate, the destination pattern of the conflicts matters in addition to their number.

For Kaspersky, AVClass2, and ClarAVy, we construct a source-specific pool containing only samples with both GT and an explicit source label, fix the same five-fold split, and compare four training conditions: (1) GT labels, (2) real source labels, (3) matched-rate uniform corruption, and (4) matched-rate structured corruption. Tables~\ref{tab:matched_decomposition_andmfc} and~\ref{tab:matched_decomposition_metamamc} report the results for AndMFC and Meta-MAMC.

\begin{table*}[ht]
\centering
\caption{Matched-pool decomposition across all three noisy sources for AndMFC.}
\label{tab:matched_decomposition_andmfc}
\setlength{\tabcolsep}{6pt}
\resizebox{\textwidth}{!}{%
\begin{tabular}{llcccc}
\toprule
\textbf{Source} & \textbf{Condition} & \textbf{ACC} & \textbf{Macro-F1} & \textbf{AMI} & \textbf{Cluster-F1} \\
\midrule
\multirow{4}{*}{Kaspersky}
 & GT                   & 96.83\% $\pm$ 0.46\% & 90.14\% $\pm$ 1.94\% & 96.69\% $\pm$ 0.43\% & 97.24\% $\pm$ 0.36\% \\
 & Real source          & 44.51\% $\pm$ 0.44\% & 20.64\% $\pm$ 0.34\% & 74.01\% $\pm$ 0.61\% & 77.42\% $\pm$ 0.36\% \\
 & Synthetic Uniform    & 81.44\% $\pm$ 1.16\% & 54.54\% $\pm$ 2.91\% & 81.96\% $\pm$ 1.03\% & 84.23\% $\pm$ 0.85\% \\
 & Synthetic Structured & 54.16\% $\pm$ 1.98\% & 37.18\% $\pm$ 1.37\% & 72.74\% $\pm$ 1.03\% & 74.57\% $\pm$ 0.65\% \\
\midrule
\multirow{4}{*}{AVClass2}
 & GT                   & 96.74\% $\pm$ 0.28\% & 89.98\% $\pm$ 1.03\% & 96.75\% $\pm$ 0.42\% & 97.14\% $\pm$ 0.28\% \\
 & Real source          & 56.58\% $\pm$ 1.57\% & 32.14\% $\pm$ 1.10\% & 83.91\% $\pm$ 1.07\% & 84.66\% $\pm$ 1.05\% \\
 & Synthetic Uniform    & 86.02\% $\pm$ 0.67\% & 64.53\% $\pm$ 0.71\% & 85.95\% $\pm$ 1.06\% & 87.80\% $\pm$ 0.95\% \\
 & Synthetic Structured & 79.79\% $\pm$ 0.64\% & 50.41\% $\pm$ 1.97\% & 86.69\% $\pm$ 0.40\% & 86.75\% $\pm$ 0.45\% \\
\midrule
\multirow{4}{*}{ClarAVy}
 & GT                   & 96.81\% $\pm$ 0.37\% & 89.92\% $\pm$ 1.58\% & 96.76\% $\pm$ 0.40\% & 97.24\% $\pm$ 0.35\% \\
 & Real source          & 62.18\% $\pm$ 0.31\% & 36.69\% $\pm$ 0.64\% & 87.76\% $\pm$ 0.27\% & 88.11\% $\pm$ 0.36\% \\
 & Synthetic Uniform    & 88.05\% $\pm$ 0.56\% & 69.19\% $\pm$ 3.01\% & 87.82\% $\pm$ 0.95\% & 89.51\% $\pm$ 0.55\% \\
 & Synthetic Structured & 84.56\% $\pm$ 0.75\% & 57.85\% $\pm$ 2.49\% & 89.41\% $\pm$ 0.50\% & 89.69\% $\pm$ 0.39\% \\
\bottomrule
\end{tabular}%
}
\end{table*}

\begin{table*}[ht]
\centering
\caption{Matched-pool decomposition across all three noisy sources for Meta-MAMC.}
\label{tab:matched_decomposition_metamamc}
\setlength{\tabcolsep}{6pt}
\resizebox{\textwidth}{!}{%
\begin{tabular}{llcccc}
\toprule
\textbf{Source} & \textbf{Condition} & \textbf{ACC} & \textbf{Macro-F1} & \textbf{AMI} & \textbf{Cluster-F1} \\
\midrule
\multirow{4}{*}{Kaspersky}
 & GT                   & 97.05\% $\pm$ 0.23\% & 87.84\% $\pm$ 1.96\% & 97.29\% $\pm$ 0.32\% & 97.54\% $\pm$ 0.25\% \\
 & Real source          & 45.26\% $\pm$ 0.80\% & 19.98\% $\pm$ 0.41\% & 75.73\% $\pm$ 0.90\% & 78.65\% $\pm$ 0.66\% \\
 & Synthetic Uniform    & 89.51\% $\pm$ 0.30\% & 58.53\% $\pm$ 1.33\% & 90.52\% $\pm$ 0.33\% & 91.98\% $\pm$ 0.16\% \\
 & Synthetic Structured & 55.00\% $\pm$ 7.72\% & 34.36\% $\pm$ 2.11\% & 77.57\% $\pm$ 3.00\% & 81.98\% $\pm$ 1.35\% \\
\midrule
\multirow{4}{*}{AVClass2}
 & GT                   & 96.63\% $\pm$ 0.57\% & 86.54\% $\pm$ 1.97\% & 97.06\% $\pm$ 0.56\% & 97.28\% $\pm$ 0.46\% \\
 & Real source          & 57.37\% $\pm$ 1.25\% & 31.72\% $\pm$ 1.11\% & 85.61\% $\pm$ 1.18\% & 86.02\% $\pm$ 1.02\% \\
 & Synthetic Uniform    & 91.84\% $\pm$ 0.60\% & 68.41\% $\pm$ 2.63\% & 92.35\% $\pm$ 0.46\% & 93.49\% $\pm$ 0.43\% \\
 & Synthetic Structured & 82.85\% $\pm$ 2.35\% & 51.83\% $\pm$ 2.45\% & 90.76\% $\pm$ 0.91\% & 90.85\% $\pm$ 0.96\% \\
\midrule
\multirow{4}{*}{ClarAVy}
 & GT                   & 96.84\% $\pm$ 0.46\% & 87.26\% $\pm$ 2.17\% & 97.09\% $\pm$ 0.50\% & 97.36\% $\pm$ 0.33\% \\
 & Real source          & 62.39\% $\pm$ 0.54\% & 36.96\% $\pm$ 0.78\% & 88.82\% $\pm$ 0.31\% & 88.84\% $\pm$ 0.35\% \\
 & Synthetic Uniform    & 92.38\% $\pm$ 0.60\% & 68.32\% $\pm$ 1.96\% & 92.89\% $\pm$ 0.81\% & 93.94\% $\pm$ 0.45\% \\
 & Synthetic Structured & 87.32\% $\pm$ 2.36\% & 59.85\% $\pm$ 2.25\% & 92.28\% $\pm$ 0.40\% & 92.54\% $\pm$ 0.51\% \\
\bottomrule
\end{tabular}%
}
\end{table*}

Taken together, Tables~\ref{tab:matched_decomposition_andmfc} and~\ref{tab:matched_decomposition_metamamc} support three consistent observations. First, GT remains the strongest supervision source on every matched pool, showing that the gap observed in Section~\ref{subsec:downstream_label_sources} cannot be explained by pool restriction alone. Second, matched-rate structured corruption is consistently more harmful than matched-rate uniform corruption, especially in Macro-F1, confirming that the structure of label errors matters in addition to their proportion. Third, for every noisy source and both probes, real source labels remain substantially worse than matched structured corruption.

This remaining gap suggests that real-world AV-derived supervision is not fully characterized by pairwise confusion rates alone; taxonomy mismatch, alias inconsistency, and other label-space distortions further damage exact family naming. At the same time, AMI and Cluster-F1 degrade less severely than Macro-F1, indicating that real AV-derived supervision still preserves part of the underlying family grouping structure even when exact family names become unstable.

\subsection{Confidence-aware Filtering with ClarAVy}
\label{subsec:claravy_confidence}

The decomposition above shows that even structured synthetic noise cannot fully account for the degradation caused by real AV labels. A complementary question is whether noise can be reduced \emph{post hoc}. Among the evaluated labeling sources, only ClarAVy outputs an explicit confidence score for each family prediction, enabling confidence-aware filtering. We therefore examine whether ClarAVy confidence can narrow the gap to expert-verified supervision.

We define confidence thresholds $\tau \in \{0.4, 0.5, 0.6, 0.7\}$. For each $\tau$, we retain only samples whose ClarAVy prediction confidence is at least $\tau$, and discard families with fewer than five retained samples. On each resulting subset, we train both AndMFC and Meta-MAMC under two supervision sources, namely expert-verified GT labels and ClarAVy-derived labels, using the same five-fold stratified protocol described in Section~\ref{subsec:rq_setup}. Because GT-based and ClarAVy-based training operate on the \emph{identical} retained subset at every threshold, the performance difference between the two supervision sources at each threshold is attributable solely to label quality. By contrast, comparisons \emph{across} thresholds are descriptive because the retained sample and family pool changes with $\tau$.

As the threshold increases from 0.4 to 0.7, the retained sample count drops from 4{,}421 (100\%) to 3{,}482 (78.8\%), 2{,}662 (60.2\%), and 1{,}747 (39.5\%), respectively. Any gain from filtering must therefore be weighed against the loss of coverage.

Table~\ref{tab:claravy_conf_classification} reports the resulting classification metrics. GT-based performance remains stable across all thresholds, confirming that the retained subsets are intrinsically learnable. ClarAVy-based Macro-F1 improves from $\tau = 0.4$ to $\tau = 0.6$ for both probes, but drops again at $\tau = 0.7$, indicating that aggressive filtering removes too many rare families and hurts macro-level balance. Even at the best threshold in our setting ($\tau = 0.6$), a large naming gap remains: the Macro-F1 difference between GT-based and ClarAVy-based training is still 40.15\% for AndMFC and 37.52\% for Meta-MAMC. Confidence filtering alone therefore cannot close the gap to expert-verified supervision for exact family naming.

\begin{table}[ht]
\centering
\caption{Classification performance under confidence-aware filtering.}
\label{tab:claravy_conf_classification}
\setlength{\tabcolsep}{6pt}
\resizebox{\textwidth}{!}{%
\begin{tabular}{clccccc}
\toprule
\textbf{$\tau$} & \textbf{Method} & \textbf{Label Source} & \textbf{ACC} & \textbf{Macro-Precision} & \textbf{Macro-Recall} & \textbf{Macro-F1} \\
\midrule
\multirow{4}{*}{0.4}
& AndMFC    & GT      & 97.67\% $\pm$ 0.46\% & 93.68\% $\pm$ 2.09\% & 91.88\% $\pm$ 1.75\% & 91.97\% $\pm$ 1.98\% \\
& AndMFC    & ClarAVy & 73.87\% $\pm$ 0.73\% & 48.19\% $\pm$ 0.91\% & 46.03\% $\pm$ 1.19\% & 45.72\% $\pm$ 1.06\% \\
& Meta-MAMC & GT      & 97.15\% $\pm$ 0.08\% & 89.62\% $\pm$ 1.07\% & 89.15\% $\pm$ 1.09\% & 88.47\% $\pm$ 0.96\% \\
& Meta-MAMC & ClarAVy & 73.42\% $\pm$ 0.91\% & 45.78\% $\pm$ 0.74\% & 44.85\% $\pm$ 1.08\% & 43.80\% $\pm$ 0.91\% \\
\midrule
\multirow{4}{*}{0.5}
& AndMFC    & GT      & 97.76\% $\pm$ 0.55\% & 93.46\% $\pm$ 2.73\% & 92.58\% $\pm$ 2.50\% & 92.14\% $\pm$ 2.71\% \\
& AndMFC    & ClarAVy & 81.33\% $\pm$ 0.63\% & 51.98\% $\pm$ 1.70\% & 50.57\% $\pm$ 2.15\% & 50.02\% $\pm$ 2.08\% \\
& Meta-MAMC & GT      & 96.96\% $\pm$ 0.41\% & 86.63\% $\pm$ 2.79\% & 86.87\% $\pm$ 1.50\% & 85.72\% $\pm$ 2.00\% \\
& Meta-MAMC & ClarAVy & 80.67\% $\pm$ 0.63\% & 49.42\% $\pm$ 1.33\% & 48.77\% $\pm$ 1.73\% & 47.80\% $\pm$ 1.30\% \\
\midrule
\multirow{4}{*}{0.6}
& AndMFC    & GT      & 97.86\% $\pm$ 0.40\% & 93.83\% $\pm$ 1.74\% & 92.47\% $\pm$ 1.29\% & 92.34\% $\pm$ 1.50\% \\
& AndMFC    & ClarAVy & 85.01\% $\pm$ 0.30\% & 53.76\% $\pm$ 0.95\% & 51.92\% $\pm$ 1.07\% & 52.19\% $\pm$ 0.91\% \\
& Meta-MAMC & GT      & 97.67\% $\pm$ 0.35\% & 92.36\% $\pm$ 2.46\% & 91.44\% $\pm$ 1.76\% & 90.92\% $\pm$ 2.09\% \\
& Meta-MAMC & ClarAVy & 85.09\% $\pm$ 0.97\% & 54.38\% $\pm$ 2.11\% & 53.76\% $\pm$ 2.87\% & 53.40\% $\pm$ 2.62\% \\
\midrule
\multirow{4}{*}{0.7}
& AndMFC    & GT      & 98.40\% $\pm$ 0.67\% & 94.26\% $\pm$ 2.60\% & 93.29\% $\pm$ 3.19\% & 93.28\% $\pm$ 3.04\% \\
& AndMFC    & ClarAVy & 89.24\% $\pm$ 0.49\% & 48.48\% $\pm$ 1.64\% & 48.41\% $\pm$ 1.58\% & 48.18\% $\pm$ 1.66\% \\
& Meta-MAMC & GT      & 97.94\% $\pm$ 0.49\% & 89.11\% $\pm$ 2.34\% & 89.29\% $\pm$ 1.02\% & 88.57\% $\pm$ 1.66\% \\
& Meta-MAMC & ClarAVy & 88.78\% $\pm$ 0.68\% & 46.03\% $\pm$ 2.41\% & 46.32\% $\pm$ 1.52\% & 45.86\% $\pm$ 1.87\% \\
\bottomrule
\end{tabular}%
}
\end{table}

Table~\ref{tab:claravy_conf_clustering} reports clustering-oriented metrics under the same confidence filtering protocol. In contrast to classification metrics, AMI and Cluster-F1 converge much more rapidly as the threshold increases. By $\tau = 0.6$, the AMI gap between GT-based and ClarAVy-based training narrows to 0.47\% for AndMFC and 0.15\% for Meta-MAMC, and Cluster-F1 values become nearly indistinguishable.

\begin{table}[ht]
\centering
\caption{Clustering-oriented metrics under confidence-aware filtering.}
\label{tab:claravy_conf_clustering}
\setlength{\tabcolsep}{6pt}
\resizebox{\textwidth}{!}{%
\begin{tabular}{clccccc}
\toprule
\textbf{$\tau$} & \textbf{Method} & \textbf{Label Source} & \textbf{AMI} & \textbf{Cluster Precision} & \textbf{Cluster Recall} & \textbf{Cluster-F1} \\
\midrule
\multirow{4}{*}{0.4}
& AndMFC    & GT      & 97.25\% $\pm$ 0.59\% & 97.69\% $\pm$ 0.49\% & 98.12\% $\pm$ 0.54\% & 97.91\% $\pm$ 0.51\% \\
& AndMFC    & ClarAVy & 92.14\% $\pm$ 0.65\% & 94.75\% $\pm$ 0.66\% & 90.84\% $\pm$ 0.91\% & 92.75\% $\pm$ 0.75\% \\
& Meta-MAMC & GT      & 97.07\% $\pm$ 0.19\% & 97.26\% $\pm$ 0.05\% & 97.90\% $\pm$ 0.15\% & 97.58\% $\pm$ 0.10\% \\
& Meta-MAMC & ClarAVy & 92.07\% $\pm$ 0.47\% & 94.21\% $\pm$ 0.39\% & 91.07\% $\pm$ 0.74\% & 92.61\% $\pm$ 0.54\% \\
\midrule
\multirow{4}{*}{0.5}
& AndMFC    & GT      & 97.36\% $\pm$ 0.65\% & 97.76\% $\pm$ 0.55\% & 98.25\% $\pm$ 0.42\% & 98.00\% $\pm$ 0.46\% \\
& AndMFC    & ClarAVy & 95.37\% $\pm$ 0.80\% & 96.38\% $\pm$ 0.57\% & 95.86\% $\pm$ 0.56\% & 96.12\% $\pm$ 0.55\% \\
& Meta-MAMC & GT      & 96.99\% $\pm$ 0.52\% & 97.07\% $\pm$ 0.47\% & 98.08\% $\pm$ 0.35\% & 97.57\% $\pm$ 0.38\% \\
& Meta-MAMC & ClarAVy & 95.20\% $\pm$ 0.73\% & 95.49\% $\pm$ 0.74\% & 95.98\% $\pm$ 0.44\% & 95.73\% $\pm$ 0.58\% \\
\midrule
\multirow{4}{*}{0.6}
& AndMFC    & GT      & 97.02\% $\pm$ 0.54\% & 97.86\% $\pm$ 0.40\% & 98.27\% $\pm$ 0.47\% & 98.07\% $\pm$ 0.43\% \\
& AndMFC    & ClarAVy & 96.56\% $\pm$ 0.58\% & 97.67\% $\pm$ 0.40\% & 97.78\% $\pm$ 0.51\% & 97.73\% $\pm$ 0.40\% \\
& Meta-MAMC & GT      & 97.26\% $\pm$ 0.52\% & 97.71\% $\pm$ 0.36\% & 98.16\% $\pm$ 0.14\% & 97.93\% $\pm$ 0.25\% \\
& Meta-MAMC & ClarAVy & 97.11\% $\pm$ 0.74\% & 97.26\% $\pm$ 0.65\% & 98.08\% $\pm$ 0.36\% & 97.67\% $\pm$ 0.45\% \\
\midrule
\multirow{4}{*}{0.7}
& AndMFC    & GT      & 97.77\% $\pm$ 1.07\% & 98.40\% $\pm$ 0.67\% & 98.80\% $\pm$ 0.64\% & 98.60\% $\pm$ 0.64\% \\
& AndMFC    & ClarAVy & 97.48\% $\pm$ 1.02\% & 98.05\% $\pm$ 0.58\% & 98.74\% $\pm$ 0.67\% & 98.40\% $\pm$ 0.59\% \\
& Meta-MAMC & GT      & 97.79\% $\pm$ 0.51\% & 98.11\% $\pm$ 0.50\% & 98.86\% $\pm$ 0.40\% & 98.48\% $\pm$ 0.43\% \\
& Meta-MAMC & ClarAVy & 97.12\% $\pm$ 0.53\% & 97.37\% $\pm$ 0.55\% & 98.63\% $\pm$ 0.49\% & 97.99\% $\pm$ 0.47\% \\
\bottomrule
\end{tabular}%
}
\end{table}

Comparing Table~\ref{tab:claravy_conf_classification} and Table~\ref{tab:claravy_conf_clustering} reveals a clear divergence between naming accuracy and grouping quality. Even at $\tau = 0.6$, ClarAVy-based Macro F1 remains 37 to 40 points below GT-based training, whereas AMI and Cluster-F1 are nearly matched. This indicates that high confidence ClarAVy labels preserve family grouping structure much better than exact family names.

Within our experimental setting, a ClarAVy confidence threshold around $\tau = 0.6$ provides a practical trade-off, retaining about 60\% of samples while maximizing classification performance of tool derived labels and nearly closing the AMI gap. We emphasize that this is an empirical recommendation rather than a universal optimum, because the retained pool changes with $\tau$. For grouping oriented tasks such as clustering based threat triage or behavioral similarity analysis, unfiltered or lightly filtered ClarAVy labels may already be usable. For tasks that require exact family naming, such as attribution, lineage tracking, or per family detection rate auditing, expert verified labels remain necessary. Beyond the main downstream evaluation, Appendix~\ref{app:temporal_split} and Appendix~\ref{app:zeroday_analysis} further examine temporal split and zero-day family settings, showing that AndroTruth also supports the study of concept drift and novel family behavior beyond standard IID classification.

\section{Discussion}
The central implication of this work is that label quality is not a peripheral annotation issue, but a first-order determinant of empirical conclusions in Android malware family research. The consistency of the supervision-source ranking across two static-feature-based probes strengthens the conclusion that supervision quality has a larger effect on downstream reliability than probe choice.

From a practical perspective, the confidence-aware filtering analysis in Section~\ref{subsec:claravy_confidence} reveals an important asymmetry between grouping quality and naming accuracy. At $\tau = 0.6$, AMI nearly matches expert-verified supervision, yet the Macro-F1 gap remains over 37 percentage points. Confidence filtering should therefore be viewed as a noise-reduction strategy for grouping-oriented tasks rather than as a replacement for expert-verified labels when precise family identification is required. The matched-pool decomposition in Section~\ref{subsec:matched_pool_decomposition} further reveals that the structure of label errors matters beyond their rate, with structured corruption consistently more harmful than uniform corruption. Real AV-derived labels perform even worse, suggesting that additional distortions such as taxonomy mismatch and alias inconsistency are not fully captured by pairwise confusion rates alone.

Our current downstream evaluation is limited to two open-source frameworks operating on Drebin-style static features. Whether the same magnitude of degradation holds for classifiers based on dynamic behavior, graph representations, or deep embeddings remains an open question. Nevertheless, because the mechanism studied here acts at the supervision level rather than the feature level, we expect the qualitative conclusion to remain relevant across broader model families. Supplementary experiments on temporal splits, zero-day family settings, and \texttt{SINGLETON} sensitivity (Appendix~A.1--A.3) further confirm the robustness of these conclusions across evaluation protocols.
\section{Threats to Validity}\label{sec:validity}

\noindent\textbf{Internal validity.}
Our dataset relies on Koodous for APK retrieval and dynamic reports. Samples can be unavailable because of takedown, access restrictions, or repository coverage, and dynamic analysis is limited by API quotas as noted in Section~\ref{subsec:static_dynamic}. The human-in-the-loop curation process can also introduce transcription, alias-reconciliation, or adjudication errors. Sample-level hashes, attributable report provenance, cross-report comparison, and exclusion of unresolved conflicts reduce but do not eliminate these risks.

\noindent\textbf{Construct validity.}
Malware-family identity is not a universally fixed construct. Vendors and analysts can use different aliases, lineage boundaries, or levels of granularity, and expert reports may cite or be influenced by the same antivirus tooling evaluated in this paper. Traceability therefore does not guarantee independence, and agreement between reports can partly reflect shared tools or taxonomies. Our exact-name conflict rate combines likely labeling mistakes with some unresolved taxonomy disagreement. Alias reconciliation, conservative exclusion, and clustering-oriented metrics mitigate this concern, but do not establish absolute family truth; GT is the operational expert-report reference defined in Section~\ref{dataset_construction}.

\noindent\textbf{External validity.}
AndroTruth contains 8{,}172 samples from 187 families and is smaller than mass-collected corpora built through automated VirusTotal aggregation. This deliberate reliability-versus-scale trade-off limits coverage of commodity malware, tail-family sample sizes, and statistical power for some family-level analyses. Reports from 42 vendors and analysts may also overrepresent high-profile or behaviorally distinctive campaigns, while source imbalance can create source-family confounding. Furthermore, both evaluated frameworks rely on Drebin-style static features; the magnitude of degradation may differ for dynamic behavior, function-call graphs, or deep embeddings. Expanding family and source coverage, adding longitudinal scan snapshots, and evaluating broader model families remain important future work.
\section{Conclusion}\label{sec:conclusion}

This paper presents AndroTruth, an Android malware family benchmark constructed from expert-traceable technical reports rather than antivirus-engine voting. Through label-audit analysis, we identify a misleading consensus phenomenon in which automated labeling tools agree with each other yet jointly deviate from expert ground truth, affecting over a quarter of explicitly labeled overlapping samples. Downstream evaluation further shows that label source fundamentally determines empirical outcomes, with expert-verified supervision dramatically outperforming all AV-derived alternatives. We additionally show that confidence-aware filtering can improve family grouping quality but cannot replace expert-verified supervision for exact family naming. Together, these results demonstrate that benchmark label reliability is a first-order factor in Android malware family evaluation. We release AndroTruth and its associated artifacts to support more reliable empirical research in this domain. In future work, we plan to continuously expand the dataset by incorporating newly published expert reports and emerging malware families, and to enrich sample-level metadata with additional analysis artifacts such as network traffic captures and fine-grained behavioral traces.
\subsubsection*{Acknowledgements}

This work is supported in part by the Hainan Province Science and Technology Special Fund under Grant ZDYF2024GXJS008, in part by the National Natural Science Foundation of China under Grants U22B2027, U2468204, and U25A20424, in part by the Joint Special Project of Beijing-Tianjin-Hebei Natural Science Foundation under Grant 25JJJJC0033, in part by the Tianjin Special Fund Project for High-quality Development of Manufacturing Industry under Grant 20251148, and in part by the Tianjin Science and Technology Plan Project under Grant 23YDPYGX00140.



\bibliographystyle{splncs04}
\bibliography{mybibliography}

\appendix
\newpage
\section{Supplementary Analysis of Experimental Content}
\subsection{Singleton Sensitivity Under a Matched-Sample Fair Protocol}
\label{app:singleton_sensitivity}

To test whether our main conclusions depend on excluding unresolved \texttt{SINGLETON} outputs, we conduct an additional fair comparison on a matched evaluation pool while retaining \texttt{SINGLETON} cases as a unified pseudo-label for AVClass2 and ClarAVy. The goal is not to treat \texttt{SINGLETON} as a semantically valid malware family, but to examine whether the relative ranking of supervision sources remains stable when unresolved outputs are preserved rather than discarded.

Table~\ref{tab:singleton_sensitivity} shows that the relative ordering is unchanged for both models: expert GT consistently outperforms ClarAVy, which in turn outperforms AVClass2 and Kaspersky. The absolute performance numbers closely mirror those in the main experiments (Table~\ref{tab:downstream_real_world}), confirming that retaining \texttt{SINGLETON} outputs does not overturn the central conclusion.

\begin{table*}[ht]
\centering
\caption{Singleton sensitivity under a matched-sample fair protocol. AVClass2 and ClarAVy retain \texttt{SINGLETON} as a unified pseudo-label.}
\label{tab:singleton_sensitivity}
\setlength{\tabcolsep}{6pt}
\resizebox{\textwidth}{!}{%
\begin{tabular}{llccc}
\toprule
\textbf{Method} & \textbf{Label Source} & \textbf{ACC} & \textbf{Macro-F1} & \textbf{AMI} \\
\midrule
\multirow{4}{*}{AndMFC} & GT & 96.79\% $\pm$ 0.48\% & 89.98\% $\pm$ 1.89\% & 96.66\% $\pm$ 0.45\% \\
 & Kaspersky & 44.53\% $\pm$ 0.43\% & 20.68\% $\pm$ 0.34\% & 74.05\% $\pm$ 0.63\% \\
 & AVClass2 & 53.00\% $\pm$ 0.89\% & 29.14\% $\pm$ 0.61\% & 80.08\% $\pm$ 0.32\% \\
 & ClarAVy & 59.58\% $\pm$ 0.74\% & 34.98\% $\pm$ 0.96\% & 85.25\% $\pm$ 0.59\% \\
\midrule
\multirow{4}{*}{Meta-MAMC} & GT & 96.91\% $\pm$ 0.27\% & 87.22\% $\pm$ 1.72\% & 97.34\% $\pm$ 0.27\% \\
 & Kaspersky & 44.78\% $\pm$ 0.77\% & 19.83\% $\pm$ 0.60\% & 75.20\% $\pm$ 1.38\% \\
 & AVClass2 & 53.19\% $\pm$ 0.72\% & 28.57\% $\pm$ 0.92\% & 81.78\% $\pm$ 0.33\% \\
 & ClarAVy & 60.04\% $\pm$ 0.99\% & 35.43\% $\pm$ 0.98\% & 86.78\% $\pm$ 0.30\% \\
\bottomrule
\end{tabular}
}
\end{table*}

\subsection{Temporal Split / Concept Drift}
\label{app:temporal_split}

To examine temporal robustness, we perform year-based temporal splits at 2020, 2021, and 2022. For each split year, the model is trained on earlier samples and tested on later samples from the same set of families, forming a closed-set temporal evaluation. This setting isolates concept drift under temporal shift while keeping the family label space fixed between training and testing.

Table~\ref{tab:temporal_closedset} shows that both models experience substantial degradation relative to random five-fold evaluation, confirming that time-aware evaluation is considerably more challenging. For AndMFC, ACC increases from 11.50\% at year 2020 to 29.88\% at year 2022, while Macro-F1 rises from 42.63\% to 62.10\% and AMI from 51.23\% to 64.10\%. Meta-MAMC performs poorly at earlier split years, with ACC near 3\% in 2020--2021 and only 75.60\% at year 2022, while Macro-F1 remains below 20.01\% across all split years despite AMI exceeding 63\% in 2022.

These results suggest that temporal generalization is substantially harder than IID evaluation, especially when training data are limited and when family behavior evolves over time. The gap between sample-level ACC and partition-level AMI further indicates that some family structure remains recoverable even when exact class assignment is unstable.

\begin{table*}[t]
\centering
\caption{Temporal closed-set evaluation with year-based splits. Models are trained on earlier samples and tested on temporally later samples from the same family set.}
\label{tab:temporal_closedset}
\setlength{\tabcolsep}{6pt}
\resizebox{\textwidth}{!}{%
\begin{tabular}{llcccccc}
\toprule
\textbf{Method} & \textbf{Year} & \textbf{Train / Test Samples} & \textbf{Train / Test Families} & \textbf{ACC} & \textbf{Macro-F1} & \textbf{AMI} & \textbf{Cluster-F1} \\
\midrule
\multirow{3}{*}{AndMFC} & 2020 & 565 / 791 & 28 / 28 & 11.50\% & 42.63\% & 51.23\% & 79.62\% \\
 & 2021 & 1038 / 2163 & 40 / 40 & 18.17\% & 43.43\% & 53.10\% & 75.30\% \\
 & 2022 & 1245 / 1861 & 34 / 34 & 29.88\% & 62.10\% & 64.10\% & 82.64\% \\
\midrule
\multirow{3}{*}{Meta-MAMC} & 2020 & 565 / 791 & 28 / 28 & 3.16\% & 2.39\% & 21.96\% & 75.44\% \\
 & 2021 & 1038 / 2163 & 40 / 40 & 3.01\% & 7.56\% & 53.41\% & 75.59\% \\
 & 2022 & 1245 / 1861 & 34 / 34 & 75.60\% & 20.01\% & 63.10\% & 84.04\% \\
\bottomrule
\end{tabular}
}
\end{table*}

\subsection{Zero-day Family Analysis}
\label{app:zeroday_analysis}

We further analyze a stricter setting where temporally later samples are divided into (i) \emph{known-family} samples whose families have appeared in the training period, and (ii) \emph{novel-family} samples from families absent from training. The known-family setting measures temporal transfer under family recurrence, whereas the novel-family setting probes how models behave when confronted with previously unseen malware families.

\paragraph{Known-family test.}
Table~\ref{tab:zeroday_known} reports performance on temporally later samples from known families. The results remain substantially lower than random five-fold evaluation and also lower than the temporal closed-set setting, confirming that temporal drift continues to affect family recognition even when the family space overlaps with training. For AndMFC, ACC increases from 10.37\% to 26.44\% across the three split years, while Meta-MAMC rises from 8.22\% to 32.72\%. AMI is consistently stronger than ACC and reaches 65.24\% for AndMFC and 67.06\% for Meta-MAMC at year 2022, suggesting that although exact labels are unstable, a nontrivial portion of the family partition remains recoverable.

\begin{table*}[t]
\centering
\caption{Zero-day evaluation on temporally later samples from known families. Models are trained on all samples before the split year and tested on later samples whose families also appear in the training period.}
\label{tab:zeroday_known}
\setlength{\tabcolsep}{6pt}
\resizebox{\textwidth}{!}{%
\begin{tabular}{llcccccccc}
\toprule
\multirow{2}{*}{\textbf{Method}} & \multirow{2}{*}{\textbf{Year}} & \multirow{2}{*}{\textbf{Train}} & \multicolumn{2}{c}{\textbf{Known Test}} & \multirow{2}{*}{\textbf{ACC}} & \multirow{2}{*}{\textbf{Macro-F1}} & \multirow{2}{*}{\textbf{AMI}} & \multirow{2}{*}{\textbf{Cluster-F1}} \\
\cmidrule(lr){4-5}
 &  &  & \textbf{Samples} & \textbf{Families} &  &  &  &  \\
\midrule
\multirow{3}{*}{AndMFC} & 2020 & 4{,}451 & 791 & 28 & 10.37\% & 26.04\% & 46.69\% & 58.22\% \\
 & 2021 & 5{,}090 & 2{,}163 & 40 & 16.18\% & 29.28\% & 56.41\% & 72.33\% \\
 & 2022 & 5{,}728 & 1{,}861 & 34 & 26.44\% & 34.56\% & 65.24\% & 81.16\% \\
\midrule
\multirow{3}{*}{Meta-MAMC} & 2020 & 4{,}451 & 791 & 28 & 8.22\% & 11.04\% & 42.30\% & 67.52\% \\
 & 2021 & 5{,}090 & 2{,}163 & 40 & 11.28\% & 10.12\% & 47.93\% & 64.95\% \\
 & 2022 & 5{,}728 & 1{,}861 & 34 & 32.72\% & 21.24\% & 67.06\% & 79.33\% \\
\bottomrule
\end{tabular}
}
\end{table*}

\paragraph{Novel-family test.}
For novel families, conventional closed-set metrics such as ACC and Macro-F1 are no longer meaningful, because the true labels are absent from the training label space. We therefore report partition-oriented and collapse-oriented indicators in Table~\ref{tab:zeroday_novel}, including AMI, Cluster-F1, the number of distinct known families predicted on the novel samples, and the dominant predicted-family ratio. Higher AMI and Cluster-F1 indicate that the model preserves more meaningful structure on novel families, whereas a high dominant predicted-family ratio indicates collapse into a small number of known families.

The results show that both methods still impose nontrivial partition structure on novel-family samples, with AMI ranging from 51.66\% to 55.92\% for Meta-MAMC and from 61.24\% to 65.10\% for AndMFC across the three split years. However, the dominant predicted-family ratio reveals different failure modes. For AndMFC, this ratio decreases from 41.06\% at year 2020 to 13.22\% at year 2021, indicating reduced collapse as the training period expands, before rising slightly to 17.03\% at year 2022. Meta-MAMC exhibits a more persistent tendency to absorb novel families into a smaller set of known ones, with dominant-family ratios between 23.65\% and 32.99\%.

Overall, the zero-day results confirm that temporally novel families remain challenging even when some latent structure is recoverable, further motivating the use of expert-verified benchmarks for temporal and open-set malware evaluation.

\begin{table*}[t]
\centering
\caption{Zero-day evaluation on temporally later samples from novel families absent from training. Since true labels are outside the training label space, we report partition-oriented metrics (AMI, Cluster-F1) and collapse indicators instead of ACC and Macro-F1.}
\label{tab:zeroday_novel}
\setlength{\tabcolsep}{6pt}
\resizebox{\textwidth}{!}{%
\begin{tabular}{llccccccc}
\toprule
\multirow{2}{*}{\textbf{Method}} & \multirow{2}{*}{\textbf{Year}} & \multicolumn{2}{c}{\textbf{Novel Test}} & \multicolumn{2}{c}{\textbf{Collapse Indicators}} & \multicolumn{2}{c}{\textbf{Partition Quality}} \\
\cmidrule(lr){3-4} \cmidrule(lr){5-6} \cmidrule(lr){7-8}
 &  & \textbf{Samples} & \textbf{Families} & \textbf{Predicted Families} & \textbf{Dominant Ratio} & \textbf{AMI} & \textbf{Cluster-F1} \\
\midrule
\multirow{3}{*}{AndMFC} & 2020 & 2{,}896 & 87 & 34 & 41.06\% & 62.08\% & 71.32\% \\
 & 2021 & 885 & 58 & 41 & 13.22\% & 65.10\% & 65.73\% \\
 & 2022 & 546 & 38 & 38 & 17.03\% & 61.24\% & 65.49\% \\
\midrule
\multirow{3}{*}{Meta-MAMC} & 2020 & 2{,}896 & 87 & 28 & 23.65\% & 51.66\% & 63.76\% \\
 & 2021 & 885 & 58 & 21 & 32.99\% & 51.95\% & 56.59\% \\
 & 2022 & 546 & 38 & 27 & 29.30\% & 55.92\% & 63.61\% \\
\bottomrule
\end{tabular}
}
\end{table*}

\end{document}